\documentclass[amssymb, twocolumn, nobibnotes, aps,pra]{revtex4-1}

\usepackage{graphicx,epsfig,epstopdf}
\usepackage{amsmath} 
\usepackage{bm}
\usepackage{color}

\begin{document}

\title{Time-resolved broadband Raman spectroscopies; A unified six-wave-mixing representation}

\author{Konstantin E. Dorfman}
\email{kdorfman@uci.edu}

\author{Benjamin P. Fingerhut}
\email{bfingerh@uci.edu}

\author{Shaul Mukamel}

\affiliation{Department of Chemistry, University of California, Irvine,
California 92697-2025, USA}
\date{\today}%

\begin{abstract}
Excited-state vibrational dynamics in molecules can be studied by an electronically off-resonant Raman process induced by a probe pulse with variable delay with respect to an actinic pulse. We establish the connection between several variants of the technique that involve either spontaneous or stimulated Raman detection and different pulse configurations. By using loop diagrams in the frequency domain we show that all signals can be described as six wave mixing  which depend on the same four point molecular correlation functions involving two transition dipoles and two polarizabilities and accompanied by a different gating. Simulations for the stochastic two-state-jump model illustrate the origin of the absorptive and dispersive features observed experimentally.
\end{abstract}

\maketitle


\vspace{0.5cm}





\section{Introduction}
Numerous types of time and frequency resolved Raman techniques have been developed towards monitoring excited and ground state vibrational dynamics in molecules \cite{Miz97,McCamant:JPCA:2003,Lee:JCP:2004,Fang:Nature:2009,Kukura:Science:2005,Kukura:AnnurevPhysChem:2007,Takeuchi:Science:2008,Kur12}. Multidimensional Raman techniques \cite{Zan09,biggs12,biggs13} employ a series of Raman processes to measure correlations between several coherence periods.  Here we consider a different class of techniques that start with a preparation (actinic) pulse $\mathcal{E}_p$ which brings the molecule into an excited electronic state and launches a vibrational dynamical response. After a variable time delay, a Raman process involving a pump $\mathcal{E}_2$ and a probe $\mathcal{E}_3$ generates the vibrational spectra. We compare four  Raman techniques: homodyne-detected Frequency-Resolved Spontaneous Raman Signal (FR-SPRS)  \cite{Ham94,Tat00,Kim01}, heterodyne-detected Time-Resolved Impulsive Stimulated Raman Signal (TR-ISRS) \cite{Fuj03,Kra13}, Transient Grating Impulsive Stimulated Raman Signal (TG-ISRS) \cite{Kno91,Hog96,Goo98,Xu01} and Femtosecond Stimulated Raman Signal (FSRS) \cite{Kukura:AnnurevPhysChem:2007,Rhi10,McC11,Kur12,Kov13,Pon13}. All provide 2D signals when displayed vs one time delay variable and a second frequency variable that reveals the Raman resonances. The factors that control the temporal and spectral resolutions are discussed. 

For simplicity we assume that the Raman process is electronically off-resonant. The effective interaction radiation/matter Hamiltonian then reads
\begin{align}\label{eq:H1}
H'(t)=\alpha \sum_{i,j}\mathcal{E}_i^{\dagger}(t)\mathcal{E}_j(t)+V\mathcal{E}_p^{\dagger}(t)+H.c.,
\end{align}
where $V$ is  the lowering (exciton annihilation) part of the dipole coupling with the actinic pulse and $\alpha $ is the excited state polarizability that couples parametrically pump and/or the probe Raman pulses. The sum over $i$ and $j$ represents the Raman process that excites the system with pulse $\mathcal{E}_j$ and deexcites it with $\mathcal{E}_i^{\dagger}$. We use complex pulse amplitudes. The electronically off-resonant Raman process induced by pulses $i$ and $j$ is instantaneous   and represented by $\alpha$ since by Heisenberg uncertainty the system can only spend a very short time in the intermediate state. 

The various configurations of excitation and deexcitation pulses for different Raman techniques will be specified below. All signals can be defined as a change of the energy of the electromagnetic field
\begin{align}\label{eq:Sdt}
S=\int_{-\infty}^{\infty}dt\frac{d}{dt}\langle a_i^{\dagger}(t)a_i(t)\rangle=\mathcal{I}\frac{2}{\hbar}\int_{-\infty}^{\infty}dt\langle\mathcal{E}_i^{\dagger}(t)\mathcal{E}_j(t)\alpha (t)\rangle,
\end{align}
where $a_i(a_i^{\dagger})$ is annihilation (creation) operator for the field $\mathcal{E}_i$ and the last equality follows from the Heisenberg equation of motion with Hamiltonian (\ref{eq:H1}). All considered techniques are related to the nonlinear susceptibility $\chi^{(5)}$. However we do not use the semiclassical $\chi^{(5)}$ approach but rather calculate the signal directly using a unified quantum field description starting with Eq. (\ref{eq:Sdt}). This is more direct and allows to treat stimulated and spontaneous Raman processes in the same fashion. Our picture uses a compact intuitive  diagrammatic representation which views all techniques as sums over paths  in the joint matter/field space involving six field matter modes and four-point material correlation functions. We show that all signals can be described in the frequency domain as a six wave mixing process with  the same molecular correlation function but convoluted with different detection windows. The role of pulse shaping can be readily discussed. 

 \begin{figure*}[t]
\begin{center}
\includegraphics[trim=0cm 0cm 0cm 0cm,angle=0, width=0.85\textwidth]{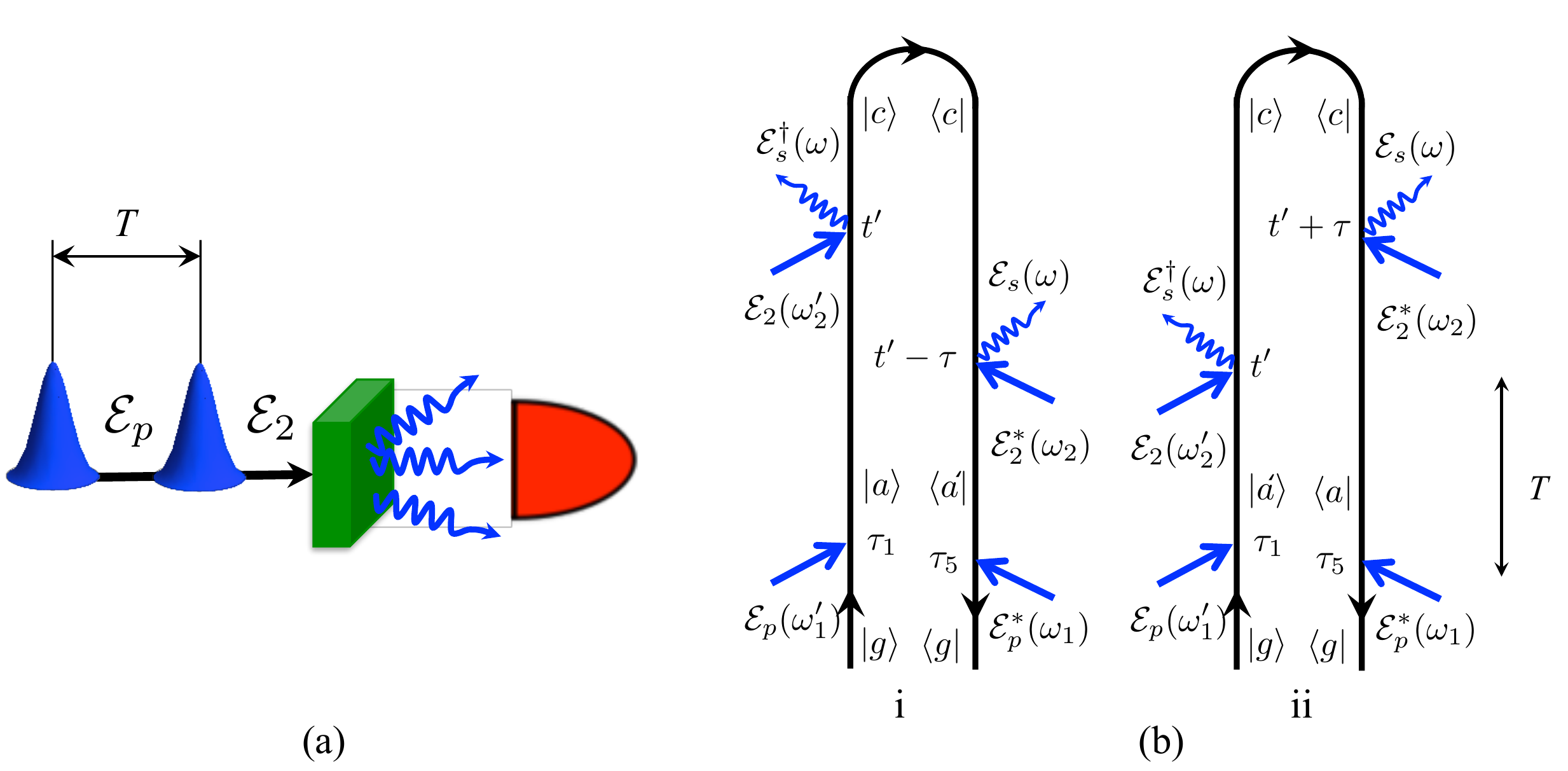}
\end{center}
\caption{(Color online) Schematic layout - (a) and loop diagrams - (b) for the FR-SPRS signal. Time translational invariance yields in this case $\omega_1'+\omega_2'-\omega+\omega-\omega_2-\omega_1=0$. Note that in contrast with the stimulated techniques where the last interaction is with the bra, here the last interaction maybe either with the bra - $i$, $\tau>0$ or with the ket - $ii$, $\tau<0$.}
\label{fig:SR}
\end{figure*}

As we show in the following, all four signals depend on two quantities that carry all the relevant information about the matter:
\begin{align}\label{eq:Fit}
 F_i(t_1,t_2,t_3)=\langle V G^{\dagger}(t_1)\alpha G^{\dagger}(t_2)\alpha G(t_3)V^{\dagger}\rangle,
\end{align}
\begin{align}\label{eq:Fiit}
F_{ii}(t_1,t_2,t_3)=\langle V G^{\dagger}(t_1)\alpha G(t_2)\alpha G(t_3)V^{\dagger}\rangle,
\end{align}
where $G(t)=(-i/\hbar)\theta(t)e^{-iHt}$ is the retarded Green's function that represents forward time evolution with the free-molecule Hamiltonian $H$. $G^{\dagger}$ represents  backward evolution. $F_i$ involves one forward and two backward evolution periods of a vibrational wave packet. $F_{ii}$ contains two forward followed by one backward propagations.

Alternatively one can express the matter correlation functions (\ref{eq:Fit}) - (\ref{eq:Fiit}) in the frequency domain

\begin{align}\label{eq:Fiw}
 F_i(\omega_1,\omega_2,\omega_3)=\langle V G^{\dagger}(\omega_1)\alpha G^{\dagger}(\omega_2)\alpha G(\omega_3)V^{\dagger}\rangle,
\end{align}
\begin{align}\label{eq:Fiiw}
F_{ii}(\omega_1,\omega_2,\omega_3)=\langle V G^{\dagger}(\omega_1)\alpha G(\omega_2)\alpha G(\omega_3)V^{\dagger}\rangle,
\end{align}
where $G(\omega)=h^{-1}/[\omega+\omega_g-H/\hbar+i\epsilon]$ and $\hbar\omega_g$ is the ground state energy. Eqs. (\ref{eq:Fit}) - (\ref{eq:Fiit}) are convenient for microscopic wave packet simulations \cite{Dor13}. Eq. (\ref{eq:Fiw}) - (\ref{eq:Fiiw}) will be used in the six-wave-mixing frequency-domain representation of the signals.

The paper is organized as follows. In Section II we present the frequency-domain expression  for the homodyne-detected FR-SPRS signal and calculate it for semiclassical bath dynamics, FSRS is discussed in Section III, TG-ISRS and TR-ISRS in Section IV. These Raman signals are compared in Section V for the stochastic two-state-jump model of line broadening. We then conclude in Section VI.

\section{The Frequency-resolved spontaneous Raman signal}

The spontaneous Raman signal is commonly measured using homodyne detection. In the FR-SPRS signal (Fig. \ref{fig:SR}) first the actinic pulse $\mathcal{E}_p$ launches the excited state dynamics. After delay period $T$, off-resonant excitation by the pump $\mathcal{E}_2$ is followed by a spontaneous emission of the Raman shifted photon. Most generally, this photon may be detected by a time-and-frequency gated detector according to Eq. (\ref{eq:S011}) and the signal is given by an overlap between detector and bare signal Wigner spectrograms. This time-and-frequency resolved detection is discussed in Appendix \ref{sec:spont} and \ref{sec:fr} and the relevant set of diagrams is shown in Fig. \ref{fig:SR1}. For a more direct comparison with the stimulated Raman techniques we shall consider here the simpler  frequency-gating. One can then eliminate two interactions with the detector and use the definition of the signal for a single mode of the radiation field Eq. (\ref{eq:Sdt}). The relevant diagrams are shown in Fig. \ref{fig:SR}. Note that in diagram $i$ the last interaction occurs with the ket, whereas in diagram $ii$ it is with the bra.

\begin{figure*}[t]
\begin{center}
\includegraphics[trim=0cm 0cm 0cm 0cm,angle=0, width=0.85\textwidth]{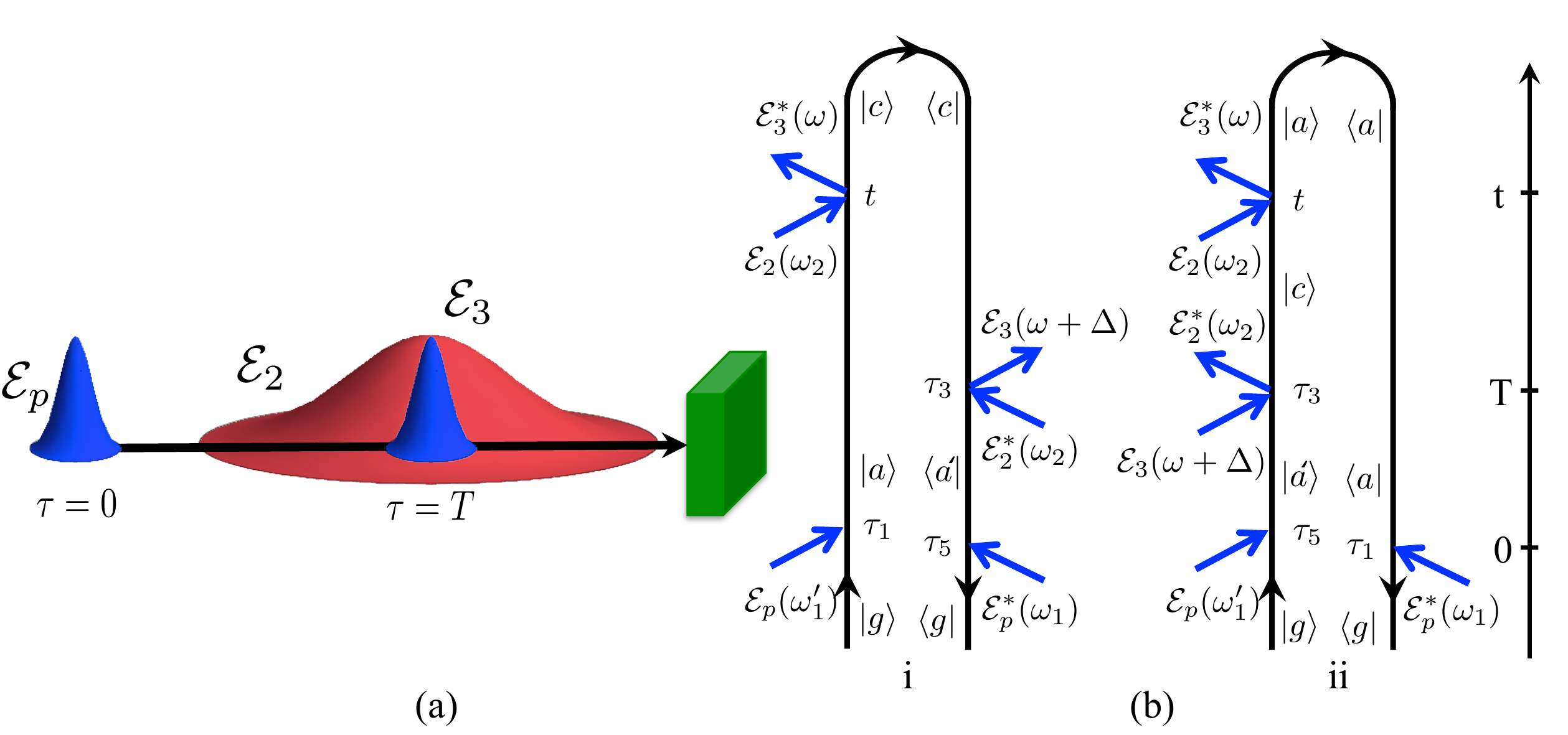}
\end{center}
\caption{(Color online) Schematic layout - (a) and loop diagrams - (b) for the FSRS signal. Time translational invariance yields $\omega_1'+\omega_2-\omega+(\omega+\Delta)-\omega_2-\omega_1=0$.}
\label{fig:FSRS}
\end{figure*}

The diagram rules\cite{Rah10} allow to write the expressions directly from the diagrams without loss of generality. To illustrate how tis works let us examine the diagram ($i$) (Fig. \ref{fig:SR}b). It can be understood using a forward and backward time evolving vibronic wave packet. First, the actinic pulse $\mathcal{E}_p(\omega_1')$ electronically excites the molecule via $V^{\dagger}$ to state $|a'\rangle$. The wavefunction then propagates  forward in time from $\tau_1$ to $t'$. Then the off-resonant pump pulse $\mathcal{E}_2(\omega_2')$ excites and instantaneous spontaneously emitted photon $\mathcal{E}_s^{\dagger}(\omega)$ deexcites the electronic transition to a different vibrational level $|c\rangle$ via $\alpha$ which then propagates backward in time from $t'$ to $t'-\tau$. The spontaneous photon $\mathcal{E}_s(\omega)$ deexcites and pump pulse $\mathcal{E}_2^{*}(\omega_2)$ excites the electronic transition from state $\langle c|$ to $\langle a|$  via $\alpha$ and the wavefunction propagates backward in time from $t'-\tau$ to $\tau_5$. The final deexcitation by pulse $\mathcal{E}_p^{*}(\omega_1)$ returns the system to its initial state by acting with $V$. Diagram $ii$ can be interpreted similarly. The diagram rules imply that the sum of the six frequencies of the various fields must be zero $\omega_1'+\omega_2'-\omega+\omega-\omega_2-\omega_1=0$. This reflects time translational invariance. For an ideal frequency-resolved detection $F_f(\omega,\bar{\omega})=\delta(\omega-\bar{\omega})$,  and Eq. (\ref{eq:Sr1}) - (\ref{eq:Sr2}) yield
\begin{align}\label{eq:spi}
&S_{FR-SPRS}(\bar{\omega},T)=-i\hbar\int_{-\infty}^{\infty}\frac{d\omega_1}{2\pi}\frac{d\omega_1'}{2\pi}\frac{d\omega_2}{2\pi}\notag\\
&\times\mathcal{D}^2(\bar{\omega})\mathcal{E}_2^{*}(\omega_2)\mathcal{E}_2(\omega_1-\omega_1'+\omega_2)\mathcal{E}_p^{*}(\omega_1)\mathcal{E}_p(\omega_1')e^{i(\omega_1-\omega_1')T}\notag\\
&\times[F_i(\omega_1,\omega_1+\omega_2-\bar{\omega},\omega_1')-F_{ii}(\omega_1,\omega_1+\omega_2-\bar{\omega},\omega_1')] .
\end{align}
where $\mathcal{D}(\omega)=\omega^3/2\pi^2c^3$ is the density of states for spontaneous modes. $\mathcal{D}(\omega)$ has taken the place of the probe pulse in the stimulated techniques. The Eq. (\ref{eq:spi})  can be alternatively recast in time domain. This is done in Appendix \ref{sec:fr}.

A simplified physical picture is obtained by a semiclassical treatment of the bath. Expanding the matter correlation function in system eigenstates (see Fig. \ref{fig:SR}b) the
Green's function for the initially prepared excited state is $G_{a}(t_1,t_2)=(-i/\hbar)\theta(t_1-t_2)e^{-(i\omega_{a}+\gamma_{a})(t_1-t_2)}$
and for the final Raman-shifted excited vibrational state \cite{Dor13,Fin13}  we have $G_{c}^{\dagger}(t,\tau_3)=(-i/\hbar)\theta(t-\tau_3)e^{i\omega_a(t-\tau_3)}\exp_-\left[\frac{i}{\hbar}\int_{\tau_3}^td\tau \omega_{ac}(\tau)\right]$, where $\omega_{ac}(\tau)$ is the vibrational frequency evolving with classical trajectories. Assuming an impulsive actinic pulse $\mathcal{E}_p(t)\simeq\mathcal{E}_p\delta(t)$ we obtain
\begin{align}\label{eq:Sfrs}
S_{FR-SPRS}&(\omega,T)=\mathcal{R}\frac{\mathcal{D}^2(\omega)}{\hbar^2}|\mathcal{E}_p|^2\sum_{a,c}|\mu_{ag}|^2\alpha_{ac}^2\notag\\
&\times\int_{-\infty}^{\infty}\frac{d\omega_2}{2\pi}\frac{d\Delta}{2\pi}\mathcal{E}_2^{*}(\omega_2)\mathcal{E}_2(\omega_2+\Delta)e^{i\Delta T}\notag\\
&\times\int_{-\infty}^{\infty}dt\int_{-\infty}^{t}d\tau e^{-\gamma_a(t+\tau)} e^{i(\omega-\omega_2)(t-\tau)-i\Delta\tau}\notag\\
&\times\exp\left[-i\int_{\tau}^{t}\omega_{ac}(\tau)d\tau\right],
\end{align}
where $\mathcal{R}$ denotes the real part and $\Delta=\omega_1-\omega_1'$ defines the spectral bandwidth of the pump pulse which translates into the spectral bandwidth of the relevant matter degrees of freedom. We have chosen to consider the entire process including the actinic pulse as a single six wave mixing event. If the actinic pulse is short enough to impulsively trigger bath dynamics but long compare to the vibrational periods so that it does not have the bandwidth to create coherences $\rho_{aa'}$ in the relevant modes we can exclude it from the description of the optical process. The process can then be viewed as a four wave mixing from a non stationary state created by the actinic pump. This is a simpler, widely used picture but it only holds in a limit parameter regime. If no dynamics is initiated and the state $\rho_{aa}$ is stationary then the pulsed experiment can be viewed as many stationary Raman experiments done in parallel. The pulse duration becomes immaterial. Note, that the above argument holds for both spontaneous and stimulated signals.

\begin{figure*}[t]
\begin{center}
\includegraphics[trim=0cm 0cm 0cm 0cm,angle=0, width=0.85\textwidth]{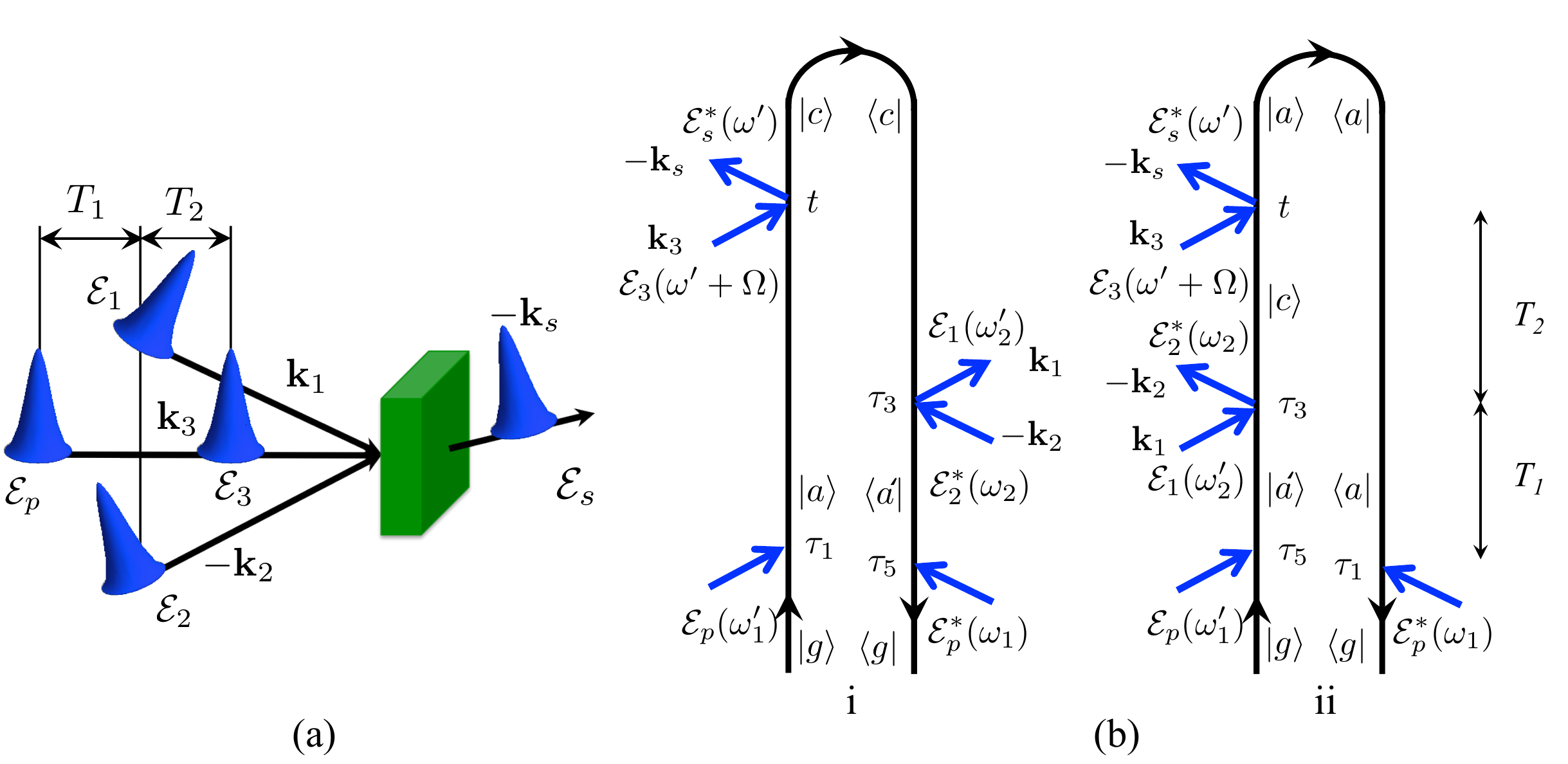}
\end{center}
\caption{(Color online) Schematic layout - (a) and loop diagrams - (b) for the TG-ISRS signal. Time translational invariance yields $\omega_1'+(\omega'+\Omega)-\omega'+\omega_2'-\omega_2-\omega_1=0$.}
\label{fig:TG}
\end{figure*}

For each pair of vibrational states $a$ and $c$,  the signal (\ref{eq:Sfrs}) can be further recast as a modulus square of a transition amplitude
\begin{align}\label{eq:Tca2}
S_{FR-SPRS}(\omega,T)=\sum_{a,c}|T_{ca}(\omega,T)|^2,
\end{align}
 where
 \begin{align}\label{eq:Tca}
 T_{ca}&(\omega,T)=\frac{\mathcal{D}(\omega)}{\hbar}\mu_{ag}^{*}\alpha_{ac}\mathcal{E}_p\int_{-\infty}^{\infty}dt\frac{d\omega_2}{2\pi}\mathcal{E}_2(\omega_2)\notag\\
 &\times e^{[i(\omega-\omega_2)-\gamma_a]t+i\omega_2 T}\exp\left[i\int_{0}^{t}\omega_{ac}(\tau)d\tau\right].
 \end{align}
As a consequence, in the limit of ideal frequency gate, the detection modes enter as frequency independent functions. Eqs. (\ref{eq:Tca2}) - (\ref{eq:Tca}) follow straightforwardly from a more general expressions Eqs. (\ref{eq:Tfi2}) - (\ref{eq:Tfi}). They show that the FR-SPRS signal is always positive and consists of purely absorptive peaks.

\begin{figure*}[t]
\begin{center}
\includegraphics[trim=0cm 0cm 0cm 0cm,angle=0, width=0.85\textwidth]{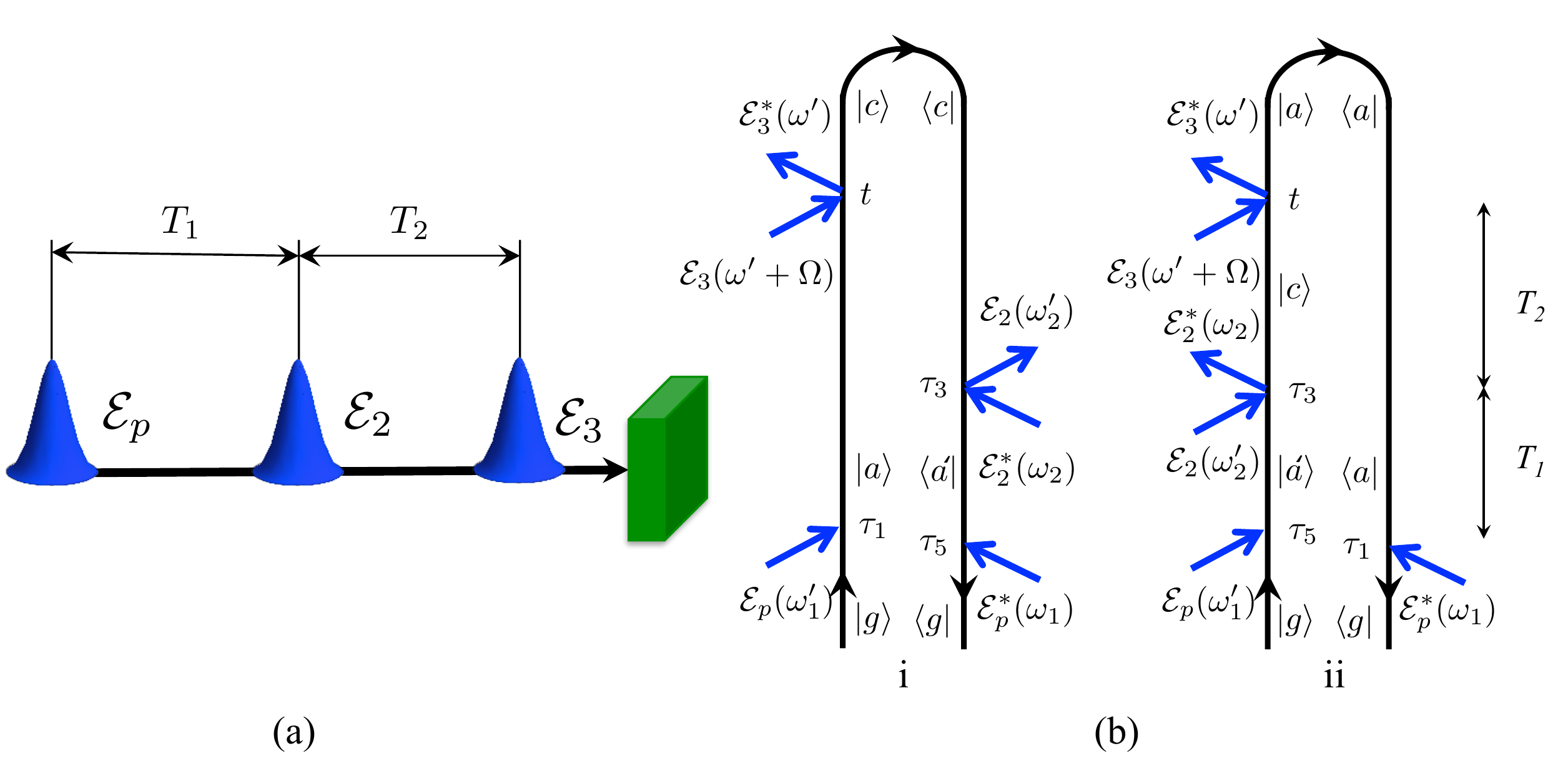}
\end{center}
\caption{(Color online) Schematic layout - (a) and loop diagrams - (b) for the TR-ISRS signal. Time translational invariance yields $\omega_1'+(\omega'+\Omega)-\omega'+\omega_2'-\omega_2-\omega_1=0$}
\label{fig:TR}
\end{figure*}

\section{The Femtosecond Stimulated Raman signal}

The stimulated signals are read off the diagrams given in Fig. \ref{fig:FSRS} as we did for FR-SPRS, a detailed derivation is given in Appendix \ref{sec:stim}, again starting with Eq. (\ref{eq:Sdt}). Following the actinic pulse $p$, after a delay $T$, pulse $2$ and the probe $3$ induce the Raman process. The signal is given by frequency dispersed probe transmission $\mathcal{E}_3(\omega)$. Diagram $i$  for FSRS can be described similar to FR-SPRS by replacing the spontaneously generated field $\mathcal{E}_s$ by a broadband probe pulse $\mathcal{E}_3$. Diagram $ii$ on the other hand is different. Following the initial electronic excitation  by actinic pulse $\mathcal{E}_p(\omega_1')$ the wavefunction  $|a'\rangle$ propagates forward in time from $\tau_5$ to $\tau_3$. At this point a Raman process involving the pump $\mathcal{E}_2^{*}(\omega_2)$ and the probe $\mathcal{E}_3(\omega+\Delta)$ promotes it to the vibrational state $|c\rangle$ and the wavefunction propagates forward in time from $\tau_3$ to $t$. After Raman deexcitation governed by $\mathcal{E}_2(\omega_2)\mathcal{E}_3^{*}(\omega)$ it then propagates backward from $t$ to $\tau_1$ in $\langle a|$ where an electronic excitation  via the actinic pulse $\mathcal{E}_p^{*}(\omega_1)$ brings the system back in its initial ground state. Assuming that pulse $2$ has a narrow bandwidth we set $\mathcal{E}_2(t-T)=\mathcal{E}_2e^{-i\omega_2(t-T)}$ and the FSRS  signal for the Raman shift $\Omega=\omega-\omega_2$ then reads\cite{Dor13}
\begin{align}\label{eq:Siwwsr1}
&S_{FSRS}(\Omega,T)=\mathcal{I}\frac{4\pi}{\hbar}\int_{-\infty}^{\infty}\frac{d\Delta}{2\pi}\frac{d\omega_1}{2\pi}\frac{d\omega_1'}{2\pi}\delta(\omega_1-\omega_1'-\Delta)\notag\\
&\times\mathcal{E}_3^{*}(\Omega+\omega_2)\mathcal{E}_3(\Omega+\omega_2+\Delta)|\mathcal{E}_2|^2\mathcal{E}_p^{*}(\omega_1)\mathcal{E}_p(\omega_1')e^{i\Delta T}\notag\\
&\times [ F_i(\omega_1,\omega_1'-\Omega,\omega_1')+F_{ii}(\omega_1,\omega_1+\Omega,\omega_1')],
\end{align}
where $\mathcal{I}$ denotes the imaginary part. In Appendix \ref{sec:fsrs} we recast Eq. (\ref{eq:Siwwsr1}) in the time domain. In contrast to FR-SPRS where the gating enters as a modulus square of an amplitude, in FSRS the symmetry between both loop branches is broken and FSRS cannot be recast as an amplitude square. While the narrowband picosecond component corresponds to $\mathcal{E}_2$ and enters as amplitude square, the femtosecond probe field $\mathcal{E}_3$ enters as $\mathcal{E}_3(\omega)\mathcal{E}_3(\omega+\Delta)$. Time translational invariance implies $\omega_1'+\omega_2-\omega+(\omega+\Delta)-\omega_2-\omega_1=0$.

Using a semiclassical description of the bath, the signal (\ref{eq:Siwwsr1}) reads
\begin{align}\label{eq:Sfss}
S_{FSRS}&(\omega-\omega_2,T)=-\mathcal{I}\frac{4}{\hbar^4}|\mathcal{E}_p|^2|\mathcal{E}_2|^2\sum_{a,c}|\mu_{ag}|^2\alpha_{ac}^2\notag\\
&\times\int_{-\infty}^{\infty}\frac{d\Delta}{2\pi}\mathcal{E}_3^{*}(\omega)\mathcal{E}_3(\omega+\Delta)e^{i\Delta T}\notag\\
&\times\int_{-\infty}^{\infty}dt\int_{-\infty}^{t}d\tau_3 e^{-\gamma_a(t+\tau_3)+i(\omega-\omega_2)(t-\tau_3)-i\Delta\tau_3}\notag\\
&\times\sin\left[\int_{\tau_3}^t\omega_{ac}(\tau)d\tau\right].
\end{align}

Here a path integral over the stochastic vibrational frequency $\omega_{ac}(t)$ determines the matter contribution to the signal.

\section{Heterodyne-detected Transient-grating and time-resolved impulsive Raman signals}

In the TG-ISRS technique, two coincident short pulses with wave vectors $\mathbf{k}_1$ and $\mathbf{k}_2$ interact with the system after a delay $T_1$ with respect to the actinic pulse and form an interference pattern with wave vector $\mathbf{k}_1-\mathbf{k}_2$. After a second delay period $T_2$, a third beam with wave vector $\mathbf{k}_3$ is scattered off the grating to generate a signal with wave vector $\mathbf{k}_s=\mathbf{k}_1-\mathbf{k}_2+\mathbf{k}_3$, which can be recorder in amplitude and phase by heterodyne detection. Traditionally TG-ISRS signals have been measured using homodyne detection of the intensity. In the heterodyne-detected TR-ISRS signals the field is mixed with the transmitted probe field. Homodyne detection suffers from various artifacts including the artificial enhancement of the modulation decay and broadening of the bandwidths in the Fourier spectra. These limitations are usually corrected by the introduction of an external local oscillator \cite{Dha94,Voe95}, which can be generated in situ by additional molecules in the same solution. Since such a molecular local oscillator contains some dynamics, the response of a combined system is not the same as from a pure optical local oscillator. The heterodyne-detected TG signals have been recently introduced to correct some of the artifacts of the homodyne detection\cite{Kra13}. In the following we compare both TG-ISRS and TR-ISRS signals using heterodyne detection.

By Fourier transforming the signal with respect to $T_2$ we obtain the vibrational spectra for different values of the first delay $T_1$. The two loop diagrams are shown in Fig. \ref{fig:TG} and the TG-ISRS is given by
\begin{align}\label{eq:TGdef}
S_{TG-ISRS}(\Omega,T_1)&=\mathcal{I}\tilde{S}_{TG-ISRS}(\Omega,T_1)\notag\\
&\times\int_{-\infty}^{\infty}\frac{d\omega'}{2\pi}\mathcal{E}_s^{*}(\omega')\mathcal{E}_3(\omega'-\Omega),
\end{align}
where $\Omega$ is the frequency conjugated to $T_2$ and
\begin{align}\label{eq:Stgi1}
\tilde{S}_{TG-ISRS}&(\Omega,T_1)=\frac{2}{\hbar}\int_{-\infty}^{\infty}\frac{d\omega_1}{2\pi}\frac{d\omega_1'}{2\pi}\frac{d\omega_2}{2\pi}\mathcal{E}_p^{*}(\omega_1)\mathcal{E}_p(\omega_1')\notag\\
&\times\mathcal{E}_2^{*}(\omega_2)\mathcal{E}_1(\omega_1-\omega_1'+\omega_2+\Omega)e^{i(\omega_1-\omega_1')T_1}\notag\\
&\times  [F_i(\omega_1,\omega_1'-\Omega,\omega_1')+F_{ii}(\omega_1,\omega_1+\Omega,\omega_1')],
\end{align}
Again, the same correlation functions $F_i$ and $F_{ii}$ as in FR-SPRS and FSRS  fully determine the matter response. Despite the fact that Eq. (\ref{eq:Stgi1}) closely resembles Eq. (\ref{eq:Siwwsr1}) the difference in detection can become crucial. In contrast to FSRS  that records the frequency dispersed transmission of the probe pulse $\mathcal{E}_3$, in TG-ISRS the signal is measured in the time domain vs two delays $T_1$, $T_2$. $T_1$ controls the time resolution, whereas $T_2$ governs the spectral resolution. Furthermore, the narrowband pump $\mathcal{E}_2$ in FSRS  enters as a modulus square $|\mathcal{E}_2|^2$, whereas a broadband fields $\mathcal{E}_1$ and $\mathcal{E}_2$ enter as $\mathcal{E}_2^{*}(\omega_2)\mathcal{E}_1(\omega_1-\omega_1'+\omega_2+\Omega)$. Instead of a single probe $\mathcal{E}_3$ in FSRS a probe pulse $\mathcal{E}_3$ is scattered off the grating and gives rise to a field $\mathcal{E}_s$. Therefore the four fields in TG-ISRS provides more control parameters for manipulating the signal compared to FSRS/FR-SPRS.

The TR-ISRS experiment uses collinear pump and probe and the probe transmission is measured. This signal is a special case of the TG-ISRS with a single pump pulse $\mathcal{E}_1=\mathcal{E}_2$ and $\mathcal{E}_s=\mathcal{E}_3$ is a probe pulse  (see Fig. \ref{fig:TR}).

\begin{align}\label{eq:TRdef}
S_{TR-ISRS}(\Omega,T_1)&=\mathcal{I}\tilde{S}_{TR-ISRS}(\Omega,T_1)\notag\\
&\times\int_{-\infty}^{\infty}\frac{d\omega'}{2\pi}\mathcal{E}_3^{*}(\omega')\mathcal{E}_3(\omega'-\Omega),
\end{align}
where 
\begin{align}\label{eq:Stri1}
\tilde{S}_{TR-ISRS}&(\Omega,T_1)=\frac{2}{\hbar}\int_{-\infty}^{\infty}\frac{d\omega_1}{2\pi}\frac{d\omega_1'}{2\pi}\frac{d\omega_2}{2\pi}\mathcal{E}_p^{*}(\omega_1)\mathcal{E}_p(\omega_1')\notag\\
&\times\mathcal{E}_2^{*}(\omega_2)\mathcal{E}_2(\omega_1-\omega_1'+\omega_2+\Omega)e^{i(\omega_1-\omega_1')T_1}\notag\\
&\times  [F_i(\omega_1,\omega_1'-\Omega,\omega_1')+F_{ii}(\omega_1,\omega_1+\Omega,\omega_1')].
\end{align}
In Appendix \ref{sec:tg} we recast Eqs. (\ref{eq:Stgi1}) and (\ref{eq:Stri1}) in the time domain. 

Even though the TR-ISRS and TG-ISRS techniques  are performed in a very different experimental configurations, they carry the same information about the matter. TG-ISRS allows for spatial selection and manipulation of the field envelopes and phases, whereas TR-ISRS depends on $|\mathcal{E}_2|^2$ and is thus independent on the phase of the pump field. Furthermore, if we compare TR-ISRS to the FSRS we note, that the main difference is that the pump pulse $\mathcal{E}_2$ in FSRS is narrowband and thus enters the signal (\ref{eq:Siwwsr1}) as $|\mathcal{E}_2|^2\delta(\omega_1-\omega_1'-\Delta)$ which is independent of its phase, whereas in the case of TR-ISRS the broadband pump yields phase dependent contribution to Eq. (\ref{eq:Stri1}): $\mathcal{E}_2^{*}(\omega_2)\mathcal{E}_2(\omega_1-\omega_1'+\omega_2+\Omega)$.

 Treating the bath semiclassically Eq. (\ref{eq:TGdef}) - Eq. (\ref{eq:Stri1}) yield
\begin{align}\label{eq:Stgs}
&S_{TG-ISRS}(\Omega,T_1)\notag\\
&=-\mathcal{I}\frac{4}{\hbar^4}|\mathcal{E}_p|^2\sum_{a,c}|\mu_{ag}|^2\alpha_{ac}^2\int_{-\infty}^{\infty}\frac{d\omega'}{2\pi}\mathcal{E}_s^{*}(\omega')\mathcal{E}_3(\omega'-\Omega)e^{-i\Omega T_1}\notag\\
&\times\int_{-\infty}^{\infty}\frac{d\omega_2}{2\pi}\frac{d\Delta}{2\pi}\mathcal{E}_2^{*}(\omega_2)\mathcal{E}_1(\omega_2+\Delta)e^{i\Delta T_1}\notag\\
&\times\int_{-\infty}^{\infty}dt\int_{-\infty}^{t}d\tau_3 e^{-\gamma_a(t+\tau_3)+i\Omega t-i\Delta\tau_3}\sin\left[\int_{\tau_3}^t\omega_{ac}(\tau)d\tau\right],
\end{align}
\begin{align}\label{eq:Strs}
&S_{TR-ISRS}(\Omega,T_1)\notag\\
&=-\mathcal{I}\frac{4}{\hbar^4}|\mathcal{E}_p|^2\sum_{a,c}|\mu_{ag}|^2\alpha_{ac}^2\int_{-\infty}^{\infty}\frac{d\omega'}{2\pi}\mathcal{E}_3^{*}(\omega')\mathcal{E}_3(\omega'-\Omega)e^{-i\Omega T_1}\notag\\
&\times\int_{-\infty}^{\infty}\frac{d\omega_2}{2\pi}\frac{d\Delta}{2\pi}\mathcal{E}_2^{*}(\omega_2)\mathcal{E}_2(\omega_2+\Delta)e^{i\Delta T_1}\notag\\
&\times\int_{-\infty}^{\infty}dt\int_{-\infty}^{t}d\tau_3 e^{-\gamma_a(t+\tau_3)+i\Omega t-i\Delta\tau_3}\sin\left[\int_{\tau_3}^t\omega_{ac}(\tau)d\tau\right].
\end{align}

\section{Comparison of Raman signals for the two-state-jump model}


The stochastic Liouville equation (SLE) \cite{Kub62,Kub63,Gam95,Tan06,San06} is a powerful tool  for  the modeling of spectral lineshapes. It assumes that the observed  quantum system  is coupled to a classical bath that undergoes stochastic dynamics; the bath affects the system but the system does not affect the bath. The SLE assumes that the bath dynamics is described by a Markovian master equation and is given by
\begin{align}\label{eq:ME}
\frac{d\rho}{dt}=\hat{\mathcal{L}}\rho(t)=-\frac{i}{\hbar}[H,\rho(t)]+\hat{L}\rho(t).
\end{align}
The SLE does not account for system/bath entanglement but  provides a very convenient level of modeling for lineshapes. 

The two-state-jump  (TSJ) model is the simplest stochastic model for lineshapes. In this model the bath has two states which we denote ``up'' $u$ and ``down'' $d$. The system has two vibrational states $a$ and $c$ with $\omega_{ac}$ being the vibrational frequency unperturbed by the bath. The TSJ coupling to the vibrations is introduced by assuming that the vibrational frequency depends on the TSJ states:  $\omega_+\equiv\omega_{ac}+\delta$ for $u$ and $\omega_-\equiv\omega_{ac}-\delta$ for $d$. The total system plus bath density matrix has eight components $|\nu\nu's\rangle$ which represent the direct product of four Liouville space states $|\nu\nu'\rangle$, $\nu,\nu'=a,c$ and two bath spin states $s=u,d$ corresponding to ``up'' and ``down'' states. The up (down) jump rates $k_u(k_d)$ are connected by the detailed-balance relation $k_u/k_d=\exp[(\epsilon_d-\epsilon_u)/k_BT]$, where $\epsilon_d-\epsilon_u$ is the energy difference between $d$ and $u$ states and $T$ is the temperature. In the low temperature limit $k_BT\ll\epsilon_u-\epsilon_d$ and $k_u=0$. The system-bath coupling is  determined by the parameter $\delta$ which represents the shift of the vibrational frequency  when the bath is in the $u$ and $d$ state. Thus, the TSJ model at low temperature depends on two parameters - jump rate $k\equiv k_d$ and splitting $\delta$. Details of the general TSJ model are given in Appendix \ref{sec:TSJ0}. The high temperature limit is discussed in Appendix \ref{sec:TSJh}.

\begin{figure*}[t]
\begin{center}
\includegraphics[trim=0cm 0cm 0cm 0cm,angle=0, width=0.75\textwidth]{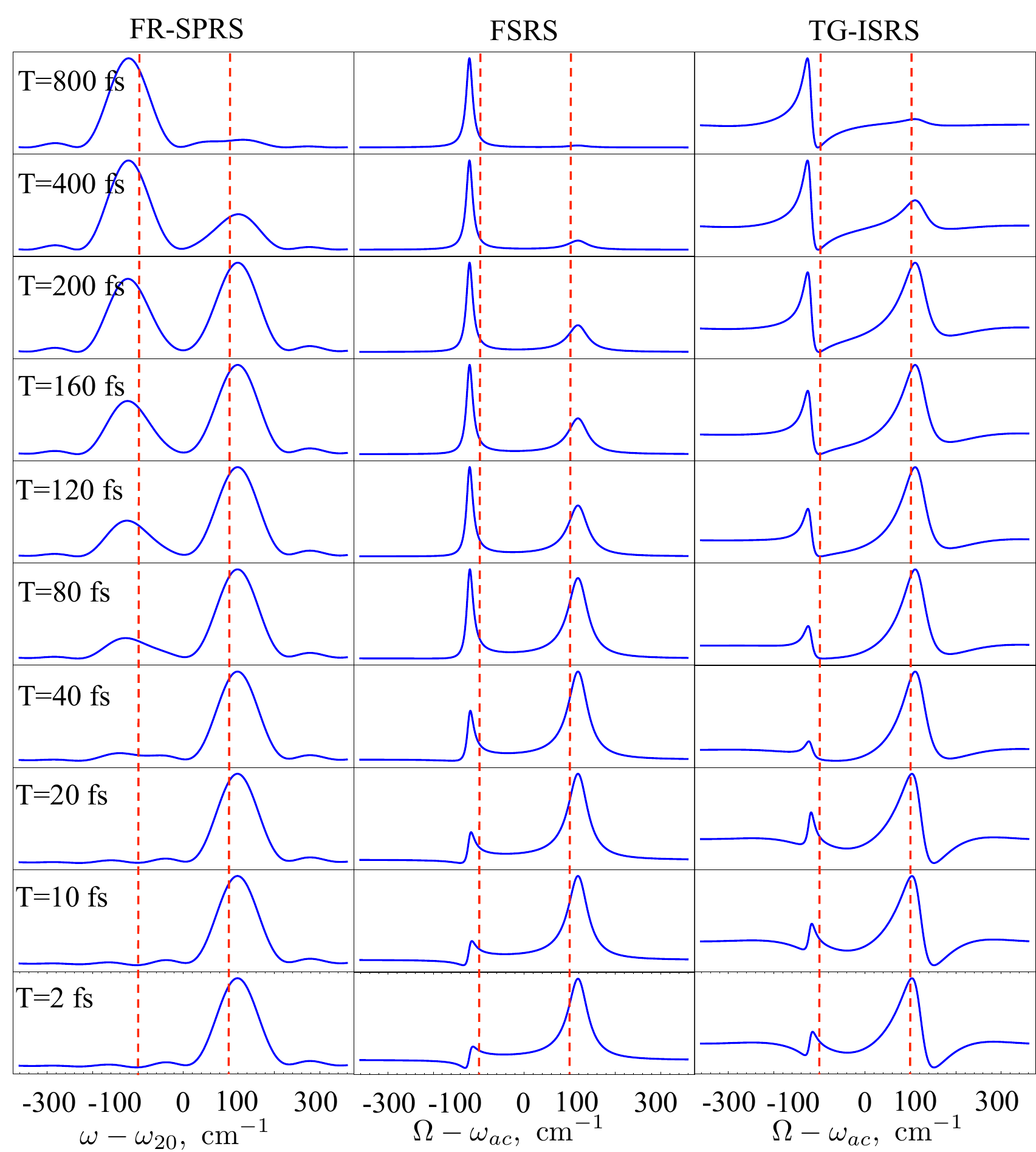}
\end{center}
\caption{(Color online) Spontaneous FR-SPRS signal in Eq. (\ref{eq:SP2}) - left column, Stimulated Raman signals - FSRS in Eq. (\ref{eq:FSRS2}) - middle and TG-ISRS (for $\mathcal{E}_2=\mathcal{E}_1$ and $\mathcal{E}_s=\mathcal{E}_3$ normalized to yield positive absorption peaks) in Eq. (\ref{eq:TG2}) - right for a TSJ model that consists of a single vibrational mode for successive values of the inter pulse delay $T$ for  $2~\text{fs}<T<800~\text{fs}$ that corresponds to a jump dynamics.  150 fs  pump $\mathcal{E}_2$ with central frequency of the pump $\omega_{20}$ for the FR-SPRS and TG-ISRS, 10 ps pump for FSRS and 5 fs probe $\mathcal{E}_3$ are centered around 800 nm wavelength. Red dashed vertical lines correspond to $\pm\delta$. The unperturbed vibrational frequency $\omega_{ac}=1000$ cm$^{-1}$. The splitting magnitude $\delta=100$ cm$^{-1}$, the inverse jump rate $k^{-1}=300$ fs, the inverse vibrational dephasing $\gamma_a^{-1}=620$ fs. If the pump pulse $\mathcal{E}_2$ in TG technique is shortened down to 5 fs TG-ISRS panel coincide with FSRS and FR-SPRS becomes broad enough to make two peak indistinguishable.}
\label{fig:sml}
\end{figure*}

We now present observables using the low temperature TSJ model. We first discuss the linear infrared absorption between the initial vibrational state $a$ and and the final state $c$ coupled to the spin. Taking the slow modulation limit (SML) ($k\ll\delta$), the linear absorption (\ref{eq:Sl}) gives a single peak with central frequency $\omega=\omega_{+}$ and width $k+\gamma_a$ where $\gamma_a$ is the dephasing rate added phenomenologically
\begin{align}\label{eq:Slsml}
S_l^{(SML)}(\omega)=\frac{2}{\hbar^2}|\mathcal{E}(\omega)|^2|\mu_{ac}|^2\frac{k+\gamma_a}{(\omega-\omega_{+})^2+(k+\gamma_a)^2}.
\end{align}
In the opposite, fast modulation (FML) -  motional narrowing limit $k\gg\delta$ Eq. (\ref{eq:Sl}) gives a peak at $\omega=\omega_{-}$ with width $\gamma_a$
\begin{align}\label{eq:Slfml}
S_l^{(FML)}(\omega)&=\frac{2}{\hbar^2}|\mathcal{E}(\omega)|^2|\mu_{ac}|^2\frac{\gamma_a}{(\omega-\omega_-)^2+\gamma_a^2}.
\end{align}
We now discuss the Raman signals for this model. The TSJ model is a special case of the semiclassical expressions given earlier. It gives simple closed expressions for the Green's functions. We first note from Eq. (\ref{eq:ME}), that unlike the signals Eqs. (\ref{eq:spi}) - (\ref{eq:Strs}) that have been written using loop diagrams in Hilbert space, the TSJ model requires a fully time ordered description in Liouville space.Each loop diagrams in Figs. \ref{fig:SR} - \ref{fig:TR} should be split into several ladder diagrams. However in the case of an impulsive actinic pulse $\mathcal{E}_p(t)=\mathcal{E}_p\delta(t)$ each loop diagram corresponds to a single ladder diagram and the total number of diagrams will remain the same. Second, the signal in Liouville space will be recast using the  matter quantity given by Eq. (\ref{eq:F12l}) where we assume that  excitation by the actinic pulse results in the state $u$.

Below we present closed expressions for the Raman signals in the SML. The general expressions for the Raman signals using TSJ model along with the derivation are given in Appendix \ref{sec:TSJ}. We further assume short pulses compared to the jump rate $k^{-1}$, dephasing $\gamma_a^{-1}$ and delay $T$: $\sigma_j\gg T^{-1}, k,\gamma_a$ where $\sigma_j$ is the spectral bandwidth of the  pump and probe pulses $j=2,3$, respectively.


The FR-SPRS signal in the SML is given by
\begin{align}\label{eq:SP2}
&S_{FR-SPRS}^{(SML)}(\omega,T)=\frac{\mathcal{D}^2(\omega)}{\hbar^2}|\mathcal{E}_p|^2\sum_{a,c}\alpha_{ac}^2|\mu_{ag}|^2e^{-2\gamma_aT}\notag\\
&\times \left(|\mathcal{E}_2(\omega-\omega_{-})|^2[1-e^{-kT}]+|\mathcal{E}_2(\omega-\omega_+)|^2e^{-kT}\right).
\end{align}


The FSRS signal for the TSJ model in the SML reads
\begin{align}\label{eq:FSRS2}
&S_{FSRS}^{(SML)}(\Omega,T)=\frac{1}{\hbar^4}|\mathcal{E}_p|^2|\mathcal{E}_2|^2|\mathcal{E}_3(\Omega+\omega_2)|^2\sum_{a,c}\alpha_{ac}^2|\mu_{ag}|^2\notag\\
&\times e^{-2\gamma_aT}\left(\frac{\gamma_a(1-e^{-kT})}{(\Omega-\omega_-)^2+\gamma_a^2}+\frac{(\gamma_a+k)e^{-kT}}{(\Omega-\omega_+)^2+(\gamma_a+k)^2}+\right.\notag\\
&\left.+\frac{k}{2\delta}e^{-kT}\left[\frac{\Omega-\omega_-}{(\Omega-\omega_-)^2+\gamma_a^2}-\frac{\Omega-\omega_+}{(\Omega-\omega_+)^2+(\gamma_a+k)^2}\right]\right)\notag\\
&-[\omega_{\pm}\leftrightarrow-\omega_{\mp}],
\end{align}
where $\Omega=\omega-\omega_2$ and the last line corresponds to the signal above with the exchanged up and down states which accounts for the anti-Stokes Raman transitions (e.g.  Stokes transition $\Omega-\omega_+$ becomes anti-Stokes $\Omega+\omega_-$ etc.)

The TG-ISRS signal in the SML yields
\begin{align}\label{eq:TG2}
&S_{TG-ISRS}^{(SML)}(\Omega,T)=\frac{1}{\hbar^4}|\mathcal{E}_p|^2\sum_{a,c}\alpha_{ac}^2|\mu_{ag}|^2\notag\\
&\times\int_{-\infty}^{\infty}\frac{d\omega'}{2\pi}\mathcal{E}_s(\omega')\mathcal{E}_3(\omega'-\Omega)\int_{-\infty}^{\infty}\frac{d\omega_2}{2\pi}\mathcal{E}_2(\omega_2)\mathcal{E}_1(\omega_2+\Omega)\notag\\
&\times e^{-2\gamma_aT}\left(\frac{\gamma_a(1-e^{-kT})}{(\Omega-\omega_-)^2+\gamma_a^2}+\frac{(\gamma_a+k)e^{-kT}}{(\Omega-\omega_+)^2+(\gamma_a+k)^2}+\right.\notag\\
&\left.+\frac{k}{2\delta}e^{-kT}\left[\frac{\Omega-\omega_-}{(\Omega-\omega_-)^2+\gamma_a^2}-\frac{\Omega-\omega_+}{(\Omega-\omega_+)^2+(\gamma_a+k)^2}\right]\right)\notag\\
&-[\omega_{\pm}\leftrightarrow-\omega_{\mp}],
\end{align}
The TR-ISRS is given by Eq. (\ref{eq:TG2}) by  simply replacing $\mathcal{E}_s\to\mathcal{E}_3$ and $\mathcal{E}_1\to\mathcal{E}_2$. 

Raman signals in the high temperature limit where $k_u=k_d=k$ are given in Appendix \ref{sec:TSJh}.


We have simulated the Raman signals (\ref{eq:SP2}) - (\ref{eq:TG2}). Short pulses provide high temporal resolution. Depending on the relation between the field and matter spectral bandwidths the pulse duration may be optimized to obtain both high temporal and spectral resolution. We assume square pulses with corresponding spectral envelope: $\mathcal{E}_j(\omega)=\mathcal{E}_j\text{sinc}\left(\frac{\omega-\omega_{j0}}{\sigma_j}\right)$, $j=2,3$. Fig. \ref{fig:sml} depicts the Raman signals for short probe pulse $\mathcal{E}_3$ compared to $\delta^{-1}$ assuming a 5 fs probe pulse $\mathcal{E}_3$, 150 fs pump pulse $\mathcal{E}_2$ and setting the jump timescale to be $k^{-1}=300$ fs, the dephasing $\gamma_a^{-1}=620$ fs and the splitting $\delta=100$ cm$^{-1}$. The left column of Fig. \ref{fig:sml} depicts the snapshot dynamics of the FR-SPRS signal. The FR-SPRS spectra consist of two broad pulse envelopes centered around $\omega_{\pm}$ according to Eq. (\ref{eq:SP2}). For small delay time $T$ the dominating peak is at $\omega_+$ corresponding to a $u$ state which exponentially decay with time. The $d$ state appears as a weak peak which grows as $1-e^{-kT}$. The FSRS (Eq. (\ref{eq:FSRS2})) is depicted in the middle column of Fig. \ref{fig:sml}. Since the probe pulse $\mathcal{E}_3$ is broadband it does not affect the shape of the spectra. Two peaks corresponding to jump frequencies $\Omega\equiv\omega-\omega_2=\omega_{\pm}$ which are visible and highlighted by red dashed lines. Depending on the delay between the actinic pulse and the probe the profile of $\omega_+$ is reduced and $\omega_-$ peak is enhanced. Note, that the vibrational coherence survives the jump since $\gamma_a<k$. At short times, the dominant contribution to the spectra is coming from the $u$ state $\Omega=\omega_+$. The linewidth is governed by a combined width of  jump rate and dephasing - $k+\gamma_a$ and is dominated by the jump rate $k$. However at long times the spectra have a single peak corresponding to a $d$ state $\Omega=\omega_-$ and its width is governed by a pure vibrational dephasing rate $\gamma_a$. Thus,  the dephasing determines the system dynamics at long times. The right column of Fig. \ref{fig:sml} shows the TG-ISRS, using the same pulse envelope for $\mathcal{E}_s(\omega)=\mathcal{E}_3(\omega)$ and $\mathcal{E}_1(\omega)=\mathcal{E}_2(\omega)$ which effectively makes it indistinguishable from the TR-ISRS. Eqs. (\ref{eq:FSRS2}) and (\ref{eq:TG2}) clearly manifest the dispersive lineshapes caused by the heterodyne detection and the broken symmetry between bra- and ket- branches of the loop.  Eq. (\ref{eq:TG2}) indicates that the TG-ISRS is very similar to the FSRS spectra if pump pulse $\mathcal{E}_2$ is broader than the splitting $\delta$. Moreover both signals coincide for a specific configuration of the pulses (see Section VI). Dynamics in the FML occurs on a very short timescale and the spectra ashow a single peak corresponding to the $d$ state (see Appendix \ref{sec:TSJ}).


\section{Discussion and Conclusions}

Eqs. (\ref{eq:spi}), (\ref{eq:Siwwsr1}), (\ref{eq:Stgi1}), and (\ref{eq:Stri1})  show that all four Raman techniques considered here depend on the same two material quantities $ F_i$ and $F_{ii}$ given in Eqs. (\ref{eq:Fit}) - (\ref{eq:Fiiw}). In the frequency domain, all techniques can be viewed as six wave mixing with constraints given by the pulse envelopes and frequency detection. In FSRS $\omega$ is fixed by the detection. Time translational invariance implies $\omega_1-\omega_1'=\Delta$. Therefore the four frequency integrations reduce to two. In FR-SPRS $\omega$ is fixed by the detection. Translational invariance $\omega_1-\omega_1'+\omega_2'-\omega_2=0$ reduces the number of integrations from four to three. Similarly in TG-ISRS and TR-ISRS, time translational invariance $\omega_1'-\omega_1+\omega_2'-\omega_2+\Omega=0$ reduces the number of integrations from four to three. Electronically-resonant Raman signals can be treated similarly but will require six point matter correlation functions since the up and down Raman transitions are no longer simultaneous. In addition resonant excitation provides a stronger signal. In this case instead of two time variables corresponding to interaction times with the Raman probe two more time variable will enter the expression for the signals. At the same time the dephasing of the electronic states is typically very short. This will keep the essential part of the formalism intact except for the information specific electronic state that is excited resonantly and enters the summation currently hidden in the polarizability. Nevertheless, our formalism can be easily extended to six-field matter interactions and the corresponding expressions can be read off the same diagrams shown in Figs. 1-4. The general six-wave mixing formalism is outside of the scope of the present paper and will be published elsewhere.

It is interesting to note that under certain conditions the FSRS  in Eq. (\ref{eq:Siwwsr1}) TG-ISRS (\ref{eq:Stgi1}) with $\mathcal{E}_s=\mathcal{E}_3$ and $\mathcal{E}_1=\mathcal{E}_2$, and TR-ISRS (\ref{eq:Stri1}) signals are identical. For example square pulses, that have been used in simulations FSRS contain $|\mathcal{E}_2|^2|\mathcal{E}_3|^2\text{sinc}^2\left(\frac{\Omega+\omega_2-\omega_{30}}{\sigma_3}\right)$. At the same time the TG-ISRS reads $|\mathcal{E}_2|^2|\mathcal{E}_3|^2\text{sinc}\left(\frac{\Omega}{\sigma_2}\right)\text{sinc}\left(\frac{\Omega}{\sigma_3}\right)$. Thus, all three techniques yield the same signal: $S_{FSRS}(\Omega,T)\simeq S_{TG-ISRS}(\Omega,T)\simeq S_{TR-ISRS}(\Omega,T)$ provided that the bandwidth of the pump and probe pulse in TG-ISRS are same as for the probe pulse in FSRS $\sigma_2=\sigma_3$ and narrowband frequency of the pump in FSRS is equal to the central frequency of the probe $\omega_2=\omega_{30}$

\begin{figure}[t]
\begin{center}
\includegraphics[trim=0cm 0cm 0cm 0cm,angle=0, width=0.49\textwidth]{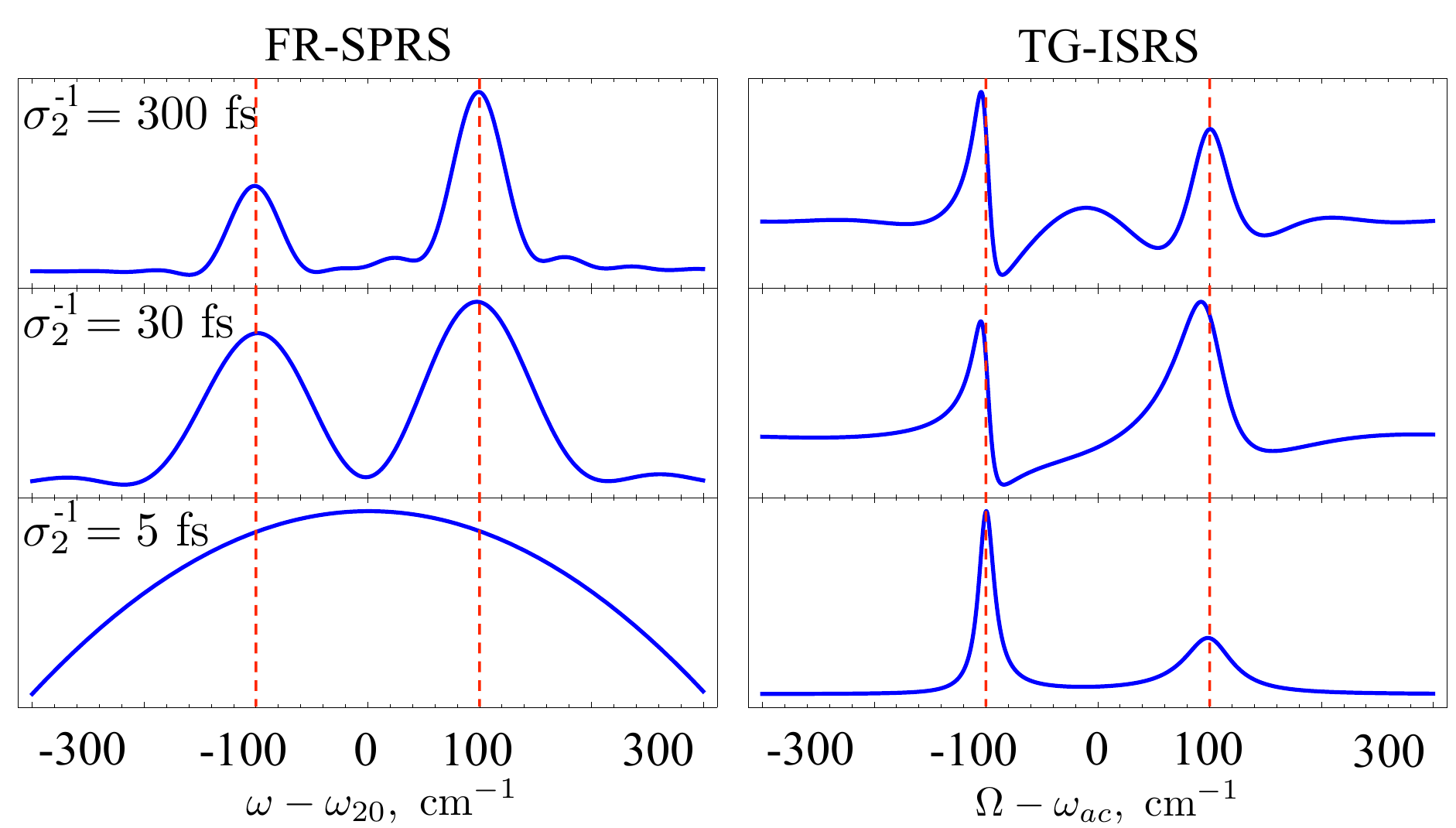}
\end{center}
\caption{(Color online) Effect of the pump pulse duration. FR-SPRS in Eq. (\ref{eq:frg}) - left and TG-ISRS in Eq. (\ref{eq:TGg}) - right in the SML for various pump pulse duration $\sigma_2$. The rest of the parameters are listed in the caption of Fig. \ref{fig:sml}.}
\label{fig:pulse}
\end{figure}

We now compare the semiclassical expressions  in Eqs. (\ref{eq:Sfrs}), (\ref{eq:Sfss}), (\ref{eq:Stgs}), (\ref{eq:Strs}). FSRS looks different from the other techniques since the signal has only two rather than three frequency integrations. This is why it has a better time resolution; the information of $ F_i$ and $F_{ii}$ is less averaged. TR-ISRS and TG-ISRS and more closely related to TR-SPRS than to FSRS. Another important point is that only the FR-SPRS signal can be recast as a modulus square of the transition amplitude. This implies that the spectra consist of purely absorptive peaks. FSRS, TG-ISRS and TR-ISRS also contain dispersive features. The underlying physical reason is clear. The spontaneous signal is dissipative where at the end of process the system is in a population state. Stimulated signals on the other hand contain diagrams that allow the system to be in a coherence state. The asymmetry between modes in ISRS signal arises since the signal is defined as a transmission of one field, whereas the dissipative signal will require measuring the transmission of all fields \cite{Rah10}. In the case of FSRS the asymmetry is caused by the frequency dispersed detection of the probe pulse which occurs only at the ket-; the bra- interacts with all the frequency components of the probe field.

We next compare the TSJ model expressions in the low temperature SML in Eqs. (\ref{eq:SP2}), (\ref{eq:FSRS2}) and (\ref{eq:TG2}). Note that in FR-SPRS  (Eq. (\ref{eq:SP2})) the spectra consist of two pump envelopes displaced by $2\delta$ as all the resonant features are contained within a pump envelope.  We next note that FSRS -  Eq. (\ref{eq:FSRS2}) does not contain any integration over the field envelopes. As shown in Sec. III, this is attributed to the frequency dispersed detection and using narrowband pump pulse overlapping with a probe. The resolution is primarily determined by the matter parameters given by $k$ - jump rate and $\gamma_a$ - vibrational dephasing. We also note that the FSRS signal contains two contributions. The first term in the square brackets in Eq. (\ref{eq:FSRS2}) corresponds to the absorptive lineshape centered around the frequencies $\omega_\pm$ dominated by a $u$ state for the short time and by a $d$ for a long time. The last term represents a dispersive lineshapes centered around the frequency $\omega_{\pm}$ which decays exponentially for both $u$ and $d$ state and is weakened by a factor of $k/2\delta$. Thus for a short time the spectra are dominated by a mixed absorptive plus dispersive peak for $u$ state with small dispersive lineshape corresponding to $d$ state.  At long times the dispersive features decay exponentially whereas the spectra contain single absorptive resonance for the $d$ state. It has been demonstrated experimentally by Kukura \cite{Kukura:Science:2005}, that at moderate times the behavior is dominated by dispersive lineshapes whereas it becomes absorptive at very long times. Similar to FSRS, TG-ISRS and TR-ISRS given by Eq. (\ref{eq:TG2}) contain both absorptive and dispersive features. Therefore all the arguments given for FSRS apply here. Furthermore we note that in contrast with FSRS, TG-ISRS contains two frequency integrals which are not coupled to the Raman resonances $\Omega=\pm\omega_{\pm}$. In addition the TG-ISRS has more freedom in tuning the pulses since it involves four pulses vs two pulses for TR-ISRS. Therefore, the resolution of the TG-ISRS and TR-ISRS are more strongly affected by the pulse envelopes than in FSRS. 

Similarly one can analyze the TSJ signals in the high temperature limit. Eqs. (\ref{eq:SP2h}), (\ref{eq:FSRS2h}) and (\ref{eq:TG2h}) show that the signals are dominated by a dispersive lineshape in the short time and are close to absorptive for the long times. In contrast to low temperatures, where the population is migrated form the $u$  state at short time to $d$ state at long times in the high temperature limit, both peaks corresponding for $u$ and $d$ state are pronounced at long times with equilibrium population in both $u$ and $d$ states.

The effect of the duration of the pump pulse $\mathcal{E}_2$ is illustrated in Fig. \ref{fig:pulse}. For a long pump pulse compared to the splitting the FR-SPRS yields two well separated peaks with the widths $\sigma_2$ corresponding to  the pulse bandwidth. For a shorter pump pulse the peaks becomes broadened and finally one cannot resolve the spectra for short enough pulse. In the case of TG-ISRS and TR-ISRS the situation is the opposite, since the effective lineshape is determined by the jump rate $k$ and dephasing $\gamma_a$. The pulse envelope results in a background which may interfere with the spectra for long enough pulses. However if the pump pulse becomes shorter than the splitting the background is uniform and the spectra contains background-free features. 

In summary we have compared four off-resonant Raman processes for studying excited-state vibrational dynamics: homodyne-detected FR-SPRS and heterodyne-detected FSRS, TG-ISRS and TR-ISRS. Using the diagram techniques we showed that all four techniques can be represented by a six field-matter interactions on the loop with the same matter information detected in a different way. Homodyne detection always yields positive signal with absorptive features determined by a pump pulse configuration which yields relatively low time and frequency resolution. Heterodyne detection yields higher temporal and spectral resolution and contain dispersive features in the spectra. Generally, heterodyne detection provides a better control of the resolution. TG-ISRS and TR-ISRS allow one to manipulate both pump and probe pulses reducing background of the signal by taking the pump short enough. In the limit of the ultrashort pump TG-ISRS/TR-ISRS yield the same resolution as the FSRS which possesses the highest resolution among all techniques utilizing overlapping narrowband pump and broadband probe pulses. The resolution of the techniques was illustrated using a two-state-jump model.

\section{Acknowledgements}

We gratefully acknowledge the support of the National Science Foundation through Grant No. CHE-1058791 and computations are supported by CHE-0840513,  the Chemical Sciences, Geosciences and Biosciences Division, Office of Basic Energy Sciences, Office of Science and US Department of Energy, National Institute of Health Grant No. GM-59230. B. P. F. gratefully acknowledges support from the Alexander-von-Humboldt Foundation through the Feodor-Lynen program.

\appendix

\section{Spontaneous signals}\label{sec:spont}

The spontaneous photon-counting signal is defined as an integrated number of photon registered by the detector:
\begin{equation}\label{eq:S0}
S(\bar{t},\bar{\omega})=\int_{-\infty}^{\infty}dt\sum_{s,s'}\langle \hat{E}_{sR}^{(tf)\dagger}(\bar{t},\bar{\omega};r_D,t)\hat{E}_{s'L}^{(tf)}(\bar{t},\bar{\omega};r_D,t)\rangle,
\end{equation}
where the angular brackets denote $\langle ...\rangle\equiv \text{Tr}[\rho(t)...]$. The density operator $\rho(t)$ is defined in the joint field-matter space of the entire system. 
Note, that Eq. (\ref{eq:S0}) represents the observable  homodyne-detected signal, and is always positive since it can be recast as a modulus square of an amplitude Eq. (\ref{eq:Tca2}). For clarity we hereafter omit the position dependence in the fields assuming that propagation between $r_G$ and $r_D$ is included in the spectral gate function $F_f$. In the case when detection is represented by a consecutive time gate $F_t$ with central time $\bar{t}$ and frequency gate $F_f$ with central frequency $\bar{\omega}$ the corresponding time-and-frequency resolved electric field reads
\begin{align}\label{eq:eft}
&E^{(tf)}(\bar{t},\bar{\omega};r_D,t)=\int_{-\infty}^{\infty}dt'F_f(t-t',\bar{\omega})F_t(t',\bar{t})\hat{E}(r_G,t'),
\end{align}
where the positive frequency part of the electric field operator is given by
\begin{equation}\label{eq:Ejs}
\hat{\mathbf{E}}(t,\mathbf{r})=\sum_{\mathbf{k}_s,\mu}\left(\frac{2\pi\hbar \omega_s}{\Omega}\right)^{1/2}\epsilon^{(\mu)}(\mathbf{k}_s)\hat{a}_{\mathbf{k}_s}e^{-i\omega_st+i\mathbf{k}_s\cdot\mathbf{r}},
\end{equation}
and $\epsilon^{(\mu)}(\mathbf{k})$ is the unit electric polarization vector of mode $(\mathbf{k}_s,\mu)$, $\mu$ being the index of polarization, $\omega_s=c|\mathbf{k}_s|$, $c$ is speed of light, $\Omega$ is quantization volume. Similarly one can apply the frequency gate first  and obtain frequency-and-time-gated field $E^{(ft)}$. Introducing the detector's Wigner spectrogram
\begin{equation}\label{eq:WD1}
W_D(\bar{t},\bar{\omega};t',\omega')=\int_{-\infty}^{\infty}\frac{d\omega}{2\pi}|F_f(\omega,\bar{\omega})|^2W_t(\bar{t};t',\omega'-\omega),
\end{equation}
where
\begin{equation}\label{eq:Wt1}
W_t(\bar{t};t',\omega)=\int_{-\infty}^{\infty}d\tau F_t^{*}(t'+\tau/2,\bar{t})F_t(t'-\tau/2,\bar{t})e^{i\omega\tau}.
\end{equation}
The detector spectrogram $W_D$ is an ordinary function of the gating time and frequency parameters which are characterized by standard deviations of the time and frequency gating $\sigma_T$ and $\sigma_{\omega}$, respectively. The structure of $W_D$ guarantees that these always satisfy the Fourier uncertainty $\sigma_\omega\sigma_T\geq 1$.
Combining Eqs. (\ref{eq:eft}) - (\ref{eq:Wt1}) we can recast Eq. (\ref{eq:S0}) in the form
\begin{equation}\label{eq:S011}
S(\bar{t},\bar{\omega})=\int_{-\infty}^{\infty}dt'\frac{d\omega'}{2\pi}W_D(\bar{t},\bar{\omega};t',\omega')W_B(t',\omega').
\end{equation}
The signal is given by the spectral and temporal overlap of a bare signal and a detector spectrogram. The bare signal contains all of the relevant information about the molecules. In order to maintain the bookkeeping of all interactions and develop a perturbative expansion for signals we adopt superoperator notation. With each ordinary operator $O$ we associate a pair of superoperators \cite{Har08} ``left'' $\hat{O}_LX=OX$, ``right'' $\hat{O}_RX=XO$, and the combination $\hat{O}_-=\hat{O}_L-\hat{O}_R$. The bare spectrogram $W_B$ in the gated photon counting signal (\ref{eq:S011})  is given in terms of superoperators as:
\begin{align}\label{eq:WE01}
W_B(t',\omega')&=\frac{\mathcal{D}^2(\omega')}{\hbar^2}\int_{-\infty}^{\infty}d\tau e^{-i\omega'\tau} \mathcal{E}_2^{*}(t'+\tau)\mathcal{E}_2(t')\notag\\
&\times\langle \mathcal{T} \alpha_{nR}(t'+\tau)\alpha_{nL}(t')e^{-\frac{i}{\hbar}\int_{-\infty}^{t'}\hat{H}_-'(T)dT}\rangle,
\end{align}
The Hamiltonian superoperator in the interaction picture under the rotating-wave approximation (RWA) is given by:
\begin{equation}\label{eq:Hq}
 \hat{H}'_{q}(t)=\int d\mathbf{r}\hat{\mathbf{E}}_{q }^{\dagger}(t,\mathbf{r})\hat{\mathbf{V}}_{q}(t,\mathbf{r})+H.c,\quad q=L,R,
 \end{equation}
where $\mathbf{V}(t,\mathbf{r})=\sum_{\alpha}\mathbf{V}^{\alpha}(t)\delta(\mathbf{r}-\mathbf{r}_{\alpha})$ is a matter operator representing the lowering (exciton annihilation) part of the dipole coupling  and $\alpha$ runs over molecules in the sample located at $\mathbf{r}_{\alpha}$. The operator $\mathcal{T}$ maintains positive time ordering of superoperators, and is a key bookkeeping device. It is defined as follows: 
 \begin{align}
\mathcal{T}\hat{E}_{q}(t_1)\hat{E}_{q'}(t_2)&=\theta(t_1-t_2)\hat{E}_{q}(t_1)\hat{E}_{q'}(t_2)\notag\\
&+\theta(t_2-t_1)\hat{E}_{q'}(t_2)\hat{E}_{q}(t_1),
\end{align}
where $\theta(t)$ is the Heaviside step function. In the absence of the frequency gate $F_f(\omega,\bar{\omega})=1$ and taking the limit of the narrow time gating $W_D(t',\omega';\bar{t},\bar{\omega})=\delta(t'-\bar{t})$ the signal (\ref{eq:S011}) reads
\begin{align}\label{eq:Tfi2}
S(\bar{t})=|T_{fi}(\bar{t})|^2,
\end{align}
where 
\begin{align}\label{eq:Tfi}
T_{fi}(t)=\frac{\mathcal{D}(\omega_{fi})}{\hbar}\mathcal{E}_2(t)\langle\mathcal{T}\alpha (t)e^{-\frac{i}{\hbar}\int_{-\infty}^tH'(T)dT}\rangle
\end{align}
is a transition amplitude, $\omega_{fi}$ is the transition frequency of the matter. Similarly in the absence of time gate $F_t(t',\bar{t})=1$ taking the limit of narrow frequency gate $W_D(t',\omega';\bar{t},\bar{\omega})=\delta(\omega'-\bar{\omega})$ the signal (\ref{eq:S011}) reads
\begin{align}\label{eq:Sidw}
S(\bar{\omega})=|T_{fi}(\bar{\omega})|^2,
\end{align}
where $T_{fi}(\omega)=\int_{-\infty}^{\infty}dte^{i\omega t}T_{fi}(t)$. Therefore in the pure time or frequency detection the signal is given by the modulus square of the transition amplitude as expected \cite{Dor12}. In this paper we use Eqs. (\ref{eq:Sidw}) for FR-SPRS but the results can be extended by using Eqs. (\ref{eq:S011}) - (\ref{eq:WE01}).

\begin{figure}[t]
\begin{center}
\includegraphics[trim=0cm 0cm 0cm 0cm,angle=0, width=0.45\textwidth]{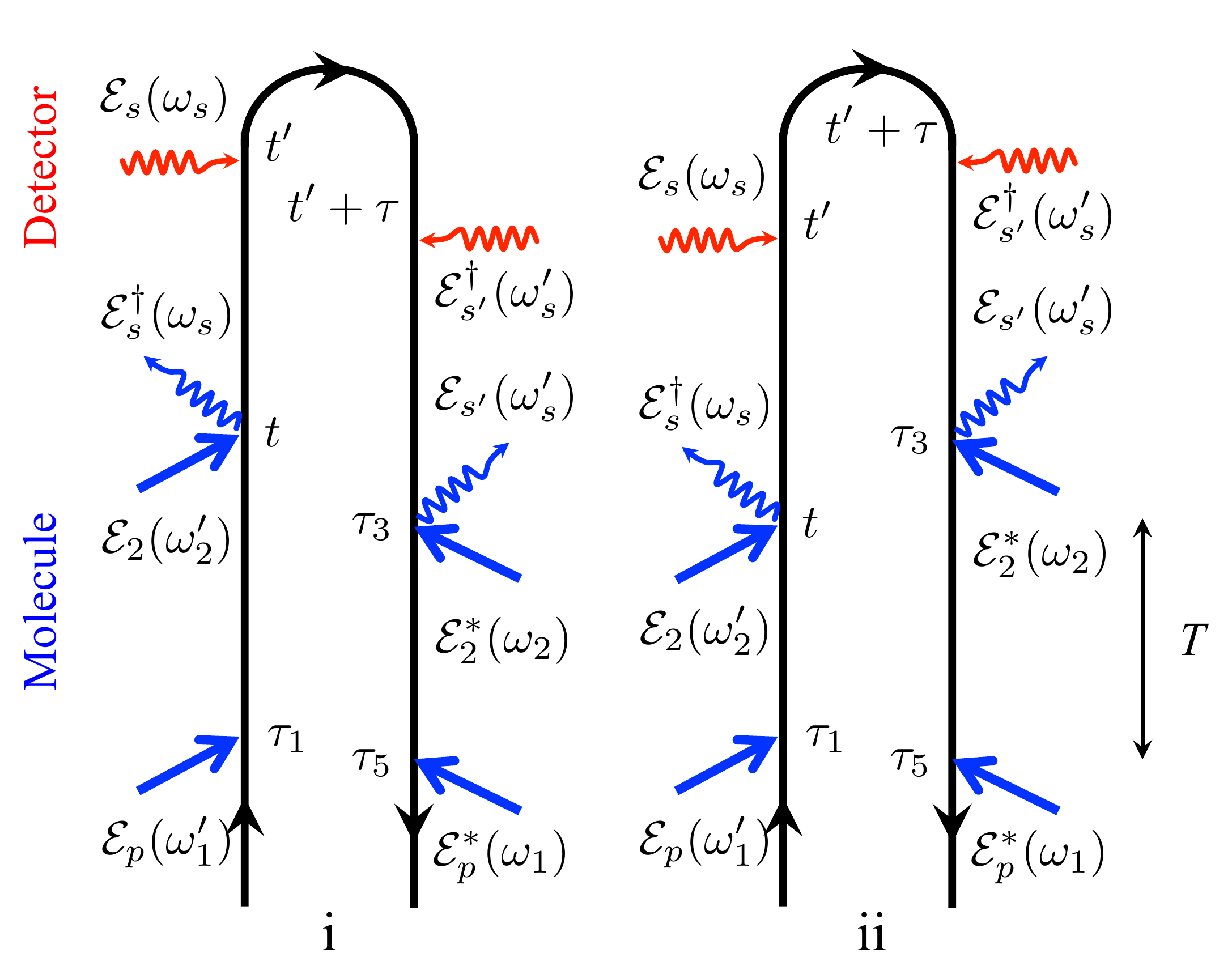}
\end{center}
\caption{(Color online) Loop diagrams for the time-and-frequency SPRS signal including interactions with the detector. Time translational invariance yields $\omega_1'+\omega_2'-\omega_s+\omega_s-\omega_s'+\omega_s'-\omega_2-\omega_1$.}
\label{fig:SR1}
\end{figure}

\section{FR-SPRS}\label{sec:fr}

We read the bare signal (see Eq. (\ref{eq:S011}) off the diagram in Fig. \ref{fig:SR1} as
\begin{align}\label{eq:Wbsri}
W_B^{(i)}&(t',\omega',T)=-i\hbar\int_0^{\infty}d\tau e^{i\omega'\tau}\int_{-\infty}^{t'}d\tau_1\int_{-\infty}^{t'-\tau}d\tau_5\notag\\
&\times\mathcal{D}^2(\bar{\omega})\mathcal{E}_2^{*}(t'-\tau-T)\mathcal{E}_2(t'-T)\mathcal{E}_p^{*}(\tau_5)\mathcal{E}_p(\tau_1)\notag\\
&\times F_i(t'-\tau-\tau_5,\tau,t'-\tau_1),
\end{align}
\begin{align}\label{eq:Wbsrii}
W_B^{(ii)}&(t',\omega',T)=i\hbar\int_{0}^{\infty}d\tau e^{-i\omega'\tau}\int_{-\infty}^{t'}d\tau_1\int_{-\infty}^{t'+\tau}d\tau_5\notag\\
&\times\mathcal{D}^2(\bar{\omega})\mathcal{E}_2^{*}(t'+\tau-T)\mathcal{E}_2(t'-T)\mathcal{E}_p^{*}(\tau_5)\mathcal{E}_p(\tau_1)\notag\\
&\times F_{ii}(t'+\tau-\tau_5,\tau,t'-\tau_1).
\end{align}
The bare signal (\ref{eq:Wbsri}) - (\ref{eq:Wbsrii}) can be alternatively recast in the frequency domain
\begin{align}
W_B^{(i)}&(t',\omega',T)=-i\hbar\int_{-\infty}^{\infty}\frac{d\omega_1}{2\pi}\frac{d\omega_1'}{2\pi}\frac{d\omega_2}{2\pi}\frac{d\omega_2'}{2\pi}\mathcal{D}^2(\bar{\omega})\notag\\
&\times\mathcal{E}_2^{*}(\omega_2)\mathcal{E}_2(\omega_2')\mathcal{E}_p^{*}(\omega_1)\mathcal{E}_p(\omega_1')e^{i(\omega_2-\omega_2'+\omega_1-\omega_1')t'}\notag\\
&\times F_i(\omega_1,\omega_1+\omega_2-\omega',\omega_1') e^{i(\omega_2'-\omega_2)T}.
\end{align}
\begin{align}
W_B^{(ii)}&(t',\omega',T)=i\hbar\int_{-\infty}^{\infty}\frac{d\omega_1}{2\pi}\frac{d\omega_1'}{2\pi}\frac{d\omega_2}{2\pi}\frac{d\omega_2'}{2\pi}\mathcal{D}^2(\bar{\omega})\notag\\
&\times\mathcal{E}_2^{*}(\omega_2)\mathcal{E}_2(\omega_2')\mathcal{E}_p^{*}(\omega_1)\mathcal{E}_p(\omega_1')e^{i(\omega_2-\omega_2'+\omega_1-\omega_1')t'}\notag\\
&\times F_{ii}(\omega_1,\omega_1+\omega_2-\omega',\omega_1') e^{i(\omega_2'-\omega_2)T},
\end{align}
To draw a full analogy with the stimulated signals we assume no time gate. In this case the detector spectrogram $W_D(\bar{t},\bar{\omega};t',\omega')=|F_f(\omega';\bar{\omega})|^2$. The time translational invariance yields
\begin{align}
\int_{-\infty}^{\infty}dt'e^{i(\omega_2-\omega_2'+\omega_1-\omega_1')t'}=2\pi\delta(\omega_2-\omega_2'+\omega_1-\omega_1')
\end{align}
which yields the signal $S_{FR-SPRS}(\bar{\omega},T)=S_{FR-SPRS}^{(i)}(\bar{\omega},T)+S_{FR-SPRS}^{(ii)}(\bar{\omega},T)$
\begin{align}\label{eq:Sr1}
&S_{FR-SPRS}^{(i)}(\bar{\omega},T)=-i\hbar\int_{-\infty}^{\infty}\frac{d\omega_1}{2\pi}\frac{d\omega_1'}{2\pi}\frac{d\omega_2}{2\pi}\frac{d\omega'}{2\pi}\mathcal{D}^2(\bar{\omega})\notag\\
&\times|F_f(\omega';\bar{\omega})|^2\mathcal{E}_2^{*}(\omega_2)\mathcal{E}_2(\omega_1-\omega_1'+\omega_2)\mathcal{E}_p^{*}(\omega_1)\mathcal{E}_p(\omega_1')\notag\\
&\times F_i(\omega_1,\omega_1+\omega_2-\omega',\omega_1') e^{i(\omega_1-\omega_1')T}.
\end{align}
\begin{align}\label{eq:Sr2}
&S_{FR-SPRS}^{(ii)}(\bar{\omega},T)=i\hbar\int_{-\infty}^{\infty}\frac{d\omega_1}{2\pi}\frac{d\omega_1'}{2\pi}\frac{d\omega_2}{2\pi}\frac{d\omega'}{2\pi}\mathcal{D}^2(\bar{\omega})\notag\\
&\times|F_f(\omega';\bar{\omega})|^2\mathcal{E}_2^{*}(\omega_2)\mathcal{E}_2(\omega_1-\omega_1'+\omega_2)\mathcal{E}_p^{*}(\omega_1)\mathcal{E}_p(\omega_1')\notag\\
&\times F_{ii}(\omega_1,\omega_1+\omega_2-\omega',\omega_1') e^{i(\omega_1-\omega_1')T}.
\end{align}
For an ideal frequency gate Eqs. (\ref{eq:Sr1}) - (\ref{eq:Sr2}) become Eq. (\ref{eq:spi}).

\section{Stimulated signals}\label{sec:stim}

Stimulated (heterodyne-detected) optical signals are defined as the energy change of the electromagnetic field
\begin{equation}
S=\int_{-\infty}^{\infty}\frac{d}{dt}\langle a^{\dagger}(t)a(t)\rangle dt,
\end{equation}
where $a(a^{\dagger})$ is annihilation (creation) operator for the field $\mathcal{E}$. The radiation-matter interaction Hamiltonian in the rotating wave approximation (RWA) is
\begin{equation}\label{eq:H1}
H'(t)=V(t)\mathcal{E}^{\dagger}(t)+H.c.,
\end{equation}
where $V(t)+V^{\dagger}(t)$ is a Heisenberg dipole operator and the electric field operator $E(t)=\mathcal{E}(t)+\mathcal{E}^{\dagger}(t)$. Both are separated into positive (non dagger) and negative (dagger) frequency components (lowering and raising photon operators, respectively).

The Heisenberg equation of motion for the field operator $E(t)$ then gives for the above integrated signal
\begin{align}\label{eq:S00}
S&=\frac{2}{\hbar}\int_{-\infty}^{\infty}dt'~\mathcal{I}\langle V(t')\mathcal{E}^{\dagger}(t')\rangle\notag\\
&=\frac{2}{\hbar}\int_{-\infty}^{\infty}\frac{d\omega'}{2\pi}\mathcal{I}\langle V(\omega')\mathcal{E}^{\dagger}(\omega')\rangle,
\end{align}
where where $\mathcal{I}$ denotes the imaginary part. The angular brackets denote $\langle...\rangle=\text{Tr}[\rho(t)...]$ with the density operator $\rho(t)$ defined in the joint field-matter space of
the entire system. In practice the temporal or spectral range of the integrations in Eq. (\ref{eq:S00}) is restricted by the response function of the detector. For a classical optical pulse one can replace the electric field operator by the expectation value $\langle\mathcal{E}\rangle=\mathcal{E}$. If the detector contains a narrow time gate with nearly $\delta$ function response $\delta(t'-t)$, Eq. (\ref{eq:S00}) yields
\begin{equation}\label{eq:St0}
S_{t}(t;\Gamma)=\frac{2}{\hbar}\mathcal{I}\mathcal{E}^{*}(t)P(t),
\end{equation}
where $P(t)=\langle V(t)\rangle$ is polarization, $\Gamma$ denotes  a set of parameters that characterize the various laser pulses. Similarly if the detector consists of a spectrometer with narrow frequency response $\delta(\omega'-\omega)$, we obtain the frequency-gated signal
\begin{equation}\label{eq:Sw0}
S_{f}(\omega;\Gamma)=\frac{2}{\hbar}\mathcal{I}\mathcal{E}^{*}(\omega)P(\omega),
\end{equation}
where $P(\omega)=\int_{-\infty}^{\infty}dtP(t)e^{i\omega t}$. Note that the two signals Eqs. (\ref{eq:St0}) and (\ref{eq:Sw0}) carry different information and are not related by a simple Fourier transform. A Wigner spectrogram representation \cite{Sto94,Dor12,Dor121} was used in \cite{Pol10} for the integrated pump probe signals Eq. (\ref{eq:S0}). Here we use loop diagrams to describe the more detailed time-  or frequency-gated signals  (\ref{eq:St0}) and (\ref{eq:Sw0}), respectively. We can also recast signals (\ref{eq:St0}) and (\ref{eq:Sw0}) in the superoperator notation for arbitrary field operator
\begin{equation}\label{eq:St}
S_{t}(t;\Gamma)=\frac{2}{\hbar}\mathcal{I}\langle \mathcal{E}_L^{\dagger}(t)V_L(t)e^{-\frac{i}{\hbar}\int_{-\infty}^{\infty}\hat{H}_-'(T)dT}\rangle,
\end{equation}
where $\Gamma$ denotes  a set of parameters that characterize the various laser pulses. Similarly if the detector consists of a spectrometer with narrow frequency response $\delta(\omega'-\omega)$, we obtain the frequency-gated signal
\begin{equation}\label{eq:Sw}
S_{f}(\omega;\Gamma)=\frac{2}{\hbar}\mathcal{I}\int_{-\infty}^{\infty}dte^{i\omega t}\langle \mathcal{E}_L^{\dagger}(\omega)V_L(t)e^{-\frac{i}{\hbar}\int_{-\infty}^{\infty}\hat{H}_-'(T)dT}\rangle.
\end{equation}
This expression is our starting point for computing the three stimulated TG-ISRS, TR-ISRS and FSRS signals.


\section{FSRS}\label{sec:fsrs}

We read the signal off the diagrams given in Fig. \ref{fig:FSRS}b

\begin{align}\label{eq:SRS}
S_{FSRS}(\omega,T)=\mathcal{I}\int_{-\infty}^{\infty}\frac{d\Delta}{2\pi}\mathcal{E}_3^{*}(\omega)\mathcal{E}_3(\omega+\Delta)\tilde{S}_{FSRS}(\omega,T;\Delta),
\end{align}
 where $\tilde{S}_{FSRS}(\omega,T;\Delta)=\tilde{S}_{FSRS}^{(i)}(\omega,T;\Delta)+\tilde{S}_{FSRS}^{(ii)}(\omega,T;\Delta)$ and
\begin{align}\label{eq:Sisr}
\tilde{S}_{FSRS}^{(i)}&(\omega,T;\Delta)=\frac{2}{\hbar}\int_{-\infty}^{\infty}dt\int_{-\infty}^td\tau_1\int_{-\infty}^td\tau_3\int_{-\infty}^{\tau_3}d\tau_5 \notag\\
&\times\mathcal{E}_2^{*}(\tau_3)\mathcal{E}_2(t)\mathcal{E}_p^{*}(\tau_5)\mathcal{E}_p(\tau_1)e^{i\omega(t-\tau_3)-i\Delta(\tau_3-T)}\notag\\
&\times  F_i(\tau_3-\tau_5,t-\tau_3,t-\tau_1),
\end{align}
\begin{align}\label{eq:Siisr}
\tilde{S}_{FSRS}^{(ii)}&(\omega,T;\Delta)=\frac{2}{\hbar}\int_{-\infty}^{\infty}dt\int_{-\infty}^td\tau_1\int_{-\infty}^td\tau_3\int_{-\infty}^{\tau_3}d\tau_5\notag\\
&\times\mathcal{E}_2^{*}(\tau_3)\mathcal{E}_2(t)\mathcal{E}_p(\tau_5)\mathcal{E}_p^{*}(\tau_1)e^{i\omega(t-\tau_3)-i\Delta(\tau_3-T)}\notag\\
&\times F_{ii}(t-\tau_1,t-\tau_3,\tau_3-\tau_5).
\end{align}
We can recast (\ref{eq:Sisr}) - (\ref{eq:Siisr}) using frequency domain matter correlation functions 
\begin{align}\label{eq:Siwwsr}
&\tilde{S}_{FSRS}^{(i)}(\omega,T;\Delta)=\mathcal{I}\frac{2}{\hbar}\int_{-\infty}^{\infty}\frac{d\omega_1}{2\pi}\frac{d\omega_1'}{2\pi}\frac{d\omega_2}{2\pi}\notag\\
&\times\mathcal{E}_2^{*}(\omega_2)\mathcal{E}_2(\omega
_2+\Delta+\omega_1'-\omega_1)\mathcal{E}_p^{*}(\omega_1)\mathcal{E}_p(\omega_1')e^{i(\omega_1-\omega_1')T}\notag\\
&\times  F_i(\omega_1,\omega_1+\omega_2-\omega-\Delta,\omega_1'),
\end{align}
\begin{align}\label{eq:Siiwwsr}
&\tilde{S}_{FSRS}^{(ii)}(\omega,T;\Delta)=\mathcal{I}\frac{2}{\hbar}\int_{-\infty}^{\infty}\frac{d\omega_1}{2\pi}\frac{d\omega_1'}{2\pi}\frac{d\omega_2}{2\pi}\notag\\
&\times\mathcal{E}_2^{*}(\omega_2)\mathcal{E}_2(\omega
_2+\Delta+\omega_1'-\omega_1)\mathcal{E}_p^{*}(\omega_1)\mathcal{E}_p(\omega_1')e^{i(\omega_1-\omega_1')T}\notag\\
&\times F_{ii}(\omega_1,\omega+\Delta-\omega_2+\omega_1',\omega_1').
\end{align}

Assuming that pulse $2$ is narrow band (picosecond) and set $\mathcal{E}_2(t-T)=\mathcal{E}_2e^{-i\omega_2(t-T)}$ the FSRS  signal for the Raman shift $\Omega=\omega-\omega_2$ then reads
\begin{align}\label{eq:SRS1}
S_{FSRS}(\Omega,T)&=\mathcal{I}\int_{-\infty}^{\infty}\frac{d\Delta}{2\pi}\mathcal{E}_3^{*}(\Omega+\omega_2)\mathcal{E}_3(\Omega+\omega_2+\Delta)\notag\\
&\times\tilde{S}_{FSRS}(\Omega,T;\Delta),
\end{align}
\begin{align}\label{eq:Sisr1}
\tilde{S}_{FSRS}^{(i)}(\Omega,T;\Delta)&=\frac{2}{\hbar}\int_{-\infty}^{\infty}dt\int_{-\infty}^td\tau_1\int_{-\infty}^td\tau_3\int_{-\infty}^{\tau_3}d\tau_5 \notag\\
&\times|\mathcal{E}_2|^2\mathcal{E}_p^{*}(\tau_5)\mathcal{E}_p(\tau_1)e^{i\Omega(t-\tau_3)-i\Delta(\tau_3-T)}\notag\\
&\times  F_i(\tau_3-\tau_5,t-\tau_3,t-\tau_1),
\end{align}
\begin{align}\label{eq:Siisr1}
\tilde{S}_{FSRS}^{(ii)}(\Omega,T;\Delta)&=\frac{2}{\hbar}\int_{-\infty}^{\infty}dt\int_{-\infty}^td\tau_1\int_{-\infty}^td\tau_3\int_{-\infty}^{\tau_3}d\tau_5\notag\\
&\times|\mathcal{E}_2|^2\mathcal{E}_p(\tau_5)\mathcal{E}_p^{*}(\tau_1)e^{i\Omega(t-\tau_3)-i\Delta(\tau_3-T)}\notag\\
&\times F_{ii}(t-\tau_1,t-\tau_3,\tau_3-\tau_5).
\end{align}
We can recast (\ref{eq:SRS1}) - (\ref{eq:Siisr1}) using frequency domain matter correlation functions and obtain Eq. (\ref{eq:Siwwsr1}).

\section{TG-ISRS and TR-ISRS}\label{sec:tg}

We read the TG-ISRS signal off the diagrams in Fig. \ref{fig:TG}b
\begin{align}\label{eq:Stgi}
\tilde{S}_{TG-ISRS}^{(i)}&(\Omega,T_1)=\frac{2}{\hbar}\int_{-\infty}^{\infty}dt\int_{-\infty}^td\tau_1\int_{-\infty}^td\tau_3\int_{-\infty}^{\tau_3}d\tau_5\notag\\
&\times\mathcal{E}_p^{*}(\tau_5)\mathcal{E}_p(\tau_1)\mathcal{E}_2^{*}(\tau_3-T)\mathcal{E}_1(\tau_3-T)e^{i\Omega(t-T_1)}\notag\\
&\times  F_i(\tau_3-\tau_5,t-\tau_3,t-\tau_1),
\end{align}
\begin{align}\label{eq:Stgii}
\tilde{S}_{TG-ISRS}^{(ii)}&(\Omega,T_1)=\frac{2}{\hbar}\int_{-\infty}^{\infty}dt\int_{-\infty}^td\tau_1\int_{-\infty}^td\tau_3\int_{-\infty}^{\tau_3}d\tau_5\notag\\
&\times\mathcal{E}_p^{*}(\tau_1)\mathcal{E}_p(\tau_5)\mathcal{E}_2^{*}(\tau_3-T)\mathcal{E}_1(\tau_3-T)e^{i\Omega(t-T_1)}\notag\\
&\times F_{ii}(t-\tau_1,t-\tau_3,\tau_3-\tau_5).
\end{align}
One can alternatively express the signals (\ref{eq:Stgi}) - (\ref{eq:Stgii}) via frequency domain correlation function of matter and obtain Eqs. (\ref{eq:Stgi1}).


Similarly we read TR-ISRS off the diagrams \ref{fig:TR}b
\begin{align}\label{eq:Stri}
\tilde{S}_{TR-ISRS}^{(i)}(\Omega,T_1)&=\frac{2}{\hbar}\int_{-\infty}^{\infty}dt\int_{-\infty}^td\tau_1\int_{-\infty}^td\tau_3\int_{-\infty}^{\tau_3}d\tau_5\notag\\
&\times\mathcal{E}_p^{*}(\tau_5)\mathcal{E}_p(\tau_1)|\mathcal{E}_2(\tau_3-T)|^2e^{i\Omega(t-T_1)}\notag\\
&\times  F_i(\tau_3-\tau_5,t-\tau_3,t-\tau_1),
\end{align}
\begin{align}\label{eq:Strii}
\tilde{S}_{TR-ISRS}^{(ii)}(\Omega,T_1)&=\frac{2}{\hbar}\int_{-\infty}^{\infty}dt\int_{-\infty}^td\tau_1\int_{-\infty}^td\tau_3\int_{-\infty}^{\tau_3}d\tau_5\notag\\
&\times\mathcal{E}_p^{*}(\tau_1)\mathcal{E}_p(\tau_5)|\mathcal{E}_2(\tau_3-T)|^2e^{i\Omega(t-T_1)}\notag\\
&\times F_{ii}(t-\tau_1,t-\tau_3,\tau_3-\tau_5).
\end{align}
Similarly one can recast Eqs. (\ref{eq:Stri}) - (\ref{eq:Strii}) in the frequency domain and obtain Eq. (\ref{eq:Stri1}).

\section{Two-state-jump model}\label{sec:TSJ0}

The Liouville operator $\hat{\mathcal{L}}$ in SLE (\ref{eq:ME}) is diagonal in the vibrational Liouville space and is thus given by four $2\times 2$ diagonal blocks in spin space \cite{San06}
\begin{align}
[\hat{\mathcal{L}}]_{\nu\nu's,\nu_1\nu_1's'}=\delta_{\nu\nu_1}\delta_{\nu'\nu_1'}[\hat{L}_S]_{s,s'}+\delta_{\nu\nu1}\delta_{\nu'\nu_1'}\delta_{ss'}[\hat{\mathcal{L}}_S]_{\nu\nu's,\nu\nu's},
\end{align}
where $\hat{L}_S$ describes the two-state-jump kinetics given by
\begin{align}
[\hat{L}_S] =
 \begin{pmatrix}
  -k_d & k_u \\
  k_d & -k_u \\
 \end{pmatrix}.
\end{align}

 The coherent part $\hat{\mathcal{L}}_S=-(i/\hbar)[H_S,...]$ vanishes for the states $|aa\rangle\rangle$, $|cc\rangle\rangle$ blocks $[\hat{\mathcal{L}}_S]_{aaaa}=[\hat{\mathcal{L}}_S]_{cccc}=0$. The remaining blocks of $\mathcal{L}_S$ reads
\begin{align}
[\hat{\mathcal{L}}_S]_{ac,ac}=-i
 \begin{pmatrix}
  \omega_{ac}+\delta & 0 \\
  0 & \omega_{ac}-\delta \\
 \end{pmatrix},
\end{align}
where $\delta$ describes the magnitude of the jump whereas $\omega_{ac}$ is the vibrational frequency unperturbed by the bath. The two Liouville space Green' functions  $\mathcal{G}(t)=-(i/\hbar)\theta(t)e^{\hat{\mathcal{L}}t}$ relevant to the Raman signal - the solution of Eq. (\ref{eq:ME}) are given by \cite{San06}
\begin{align}\label{eq:Gaa}
\mathcal{G}_{aa,aa}(t)=(-i/\hbar)\theta(t)\left[\hat{1}+\frac{1-e^{-(k_u+k_d)t}}{k_d+k_u} \begin{pmatrix}
  -k_d& k_u \\
  k_d & -k_u \\
 \end{pmatrix}\right],
 \end{align}
\begin{align}\label{eq:Gac}
\mathcal{G}_{ac,ac}(t)&=(-i/\hbar)\theta(t)\notag\\
&\times\left[\left(\frac{\eta_2}{\eta_2-\eta_1}\hat{1}-\frac{1}{\eta_2-\eta_1}\hat{\mathcal{L}}_{ac,ac}\right)e^{\eta_1t}\right.\notag\\
&\left.+\left(\frac{\eta_1}{\eta_1-\eta_2}\hat{1}-\frac{1}{\eta_1-\eta_2}\hat{\mathcal{L}}_{ac,ac}\right)e^{\eta_2t}\right],
 \end{align}
where $\hat{1}$ is unit 2$\times$2 matrix and $\eta_{1,2}=-\frac{k_d+k_u}{2}-i\omega_{ac}\pm\sqrt{\frac{(k_d+k_u)^2}{4}-\delta^2+i\delta(k_d-k_u)}$.


In the low temperature limit $k_BT\ll\epsilon_u-\epsilon_d$ and thus $k_u=0$, Eqs. (\ref{eq:Gaa}) - (\ref{eq:Gac}) are given by the following $2\times 2$ matrices
\begin{align}\label{eq:Gaal}
\mathcal{G}_{aa,aa}(t)=(-i/\hbar)\theta(t)
 \begin{pmatrix}
  e^{-kt}& 0 \\
  1-e^{-kt} & 1 \\
 \end{pmatrix},
 \end{align}
\begin{align}\label{eq:Gacl}
&\mathcal{G}_{ac,ac}(t)=(-i/\hbar)\theta(t)\notag\\
&\times\begin{pmatrix}
  e^{-(k+i\omega_{+})t}& 0 \\
  \frac{k}{k+2i\delta}[e^{-i\omega_-t}-e^{-(k+i\omega_{+})t}] &e^{-i\omega_{-}t} \\
 \end{pmatrix},
 \end{align}
where $k=k_u$ and $\omega_{\pm}=\omega_{ac}\pm\delta$. 

We now turn to a discussion of the the linear absorption between the initial vibrational state $a$ and and the final state $c$ coupled to the spin
\begin{align}\label{eq:Sl}
S_l(\omega)=\mathcal{I}\frac{2}{\hbar}|\mathcal{E}(\omega)|^2|\mu_{ac}|^2\langle\langle I_a|\mathcal{G}_{ac,ac}(\omega)|\rho_a\rangle\rangle_S,
\end{align}
where $\mathcal{G}(\omega)=\int_{-\infty}^{\infty}e^{i\omega t}\mathcal{G}(t)dt$. The initial state in the spin space is equilibrium state: $|\rho_a\rangle\rangle_S=|aa\rangle\rangle\begin{pmatrix}
  1 \\
  1 \\
 \end{pmatrix}$, and we trace over the final state $\langle\langle I|=(1,1)\text{Tr}$ where $\text{Tr}=\langle\langle aa|+\langle\langle cc|$. After matrix multiplication the time-domain matter correlation function reads
 \begin{align}
  \langle\langle I_a|\mathcal{G}_{ac,ac}(t)|\rho_a\rangle\rangle_S&=(-i/\hbar)\theta(t)\frac{2}{k+2i\delta}\notag\\
&\times\left[(k+i\delta)e^{-i\omega_{-}t}+i\delta e^{-(k+i\omega_{+})t}\right].
 \end{align}
In the SML the linear absorption (\ref{eq:Sl}) yields Eq. (\ref{eq:Slsml}) whereas in FML it gives Eq. (\ref{eq:Slfml}).

\section{Raman signals with TSJ model at low temperature}\label{sec:TSJ}

\subsection{FR-SPRS with TSJ (Low T)}

We first recast a time-domain signal Eqs. (\ref{eq:S011}), (\ref{eq:Wbsri}) - (\ref{eq:Wbsrii}) in the Liouville space breaking the loop in diagrams $i$ and $ii$ in Fig. \ref{fig:SR} for the impulsive actinic pulse $\mathcal{E}_p(t)=\mathcal{E}_p\delta(t)$:
\begin{align}\label{eq:frji}
S_{FR-SPRS}^{(i)}(\omega,&T)=-i\hbar|\mathcal{E}_p|^2\mathcal{D}^2(\omega)\int_{-\infty}^{\infty}dt'\int_0^{\infty}d\tau e^{i\omega\tau}\notag\\
&\times\mathcal{E}_2^{*}(t'-\tau-T)\mathcal{E}_2(t'-T)\mathcal{F}(\tau,t'-\tau),
\end{align}
\begin{align}\label{eq:frjii}
S_{FR-SPRS}^{(ii)}(\omega,&T)=i\hbar|\mathcal{E}_p|^2\mathcal{D}^2(\omega)\int_{-\infty}^{\infty}dt'\int_0^{\infty}d\tau e^{-i\omega\tau}\notag\\
&\times\mathcal{E}_2^{*}(t'+\tau-T)\mathcal{E}_2(t'-T)\mathcal{F}^{*}(\tau,t'),
\end{align}
where we have introduced the following matter quantity
\begin{align}\label{eq:FL}
\mathcal{F}(t_1,t_2)=\frac{i}{\hbar}\sum_{a,c}\alpha_{ac}^2|\mu_{ag}|^2\langle\langle I|\mathcal{G}_{ac,ac}(t_1)\mathcal{G}_{aa,aa}(t_2)|\rho_0\rangle\rangle_S.
\end{align}
Here we assumed that after excitation by the actinic pulse the state is $u$: $|\rho_0\rangle\rangle_S=|aa\rangle\rangle\begin{pmatrix}
  1 \\
  0 \\
 \end{pmatrix}$. Perform a  matrix multiplication using Eq. (\ref{eq:Gaal}) and (\ref{eq:Gacl}) we obtain a compact form for Eq. (\ref{eq:FL}) \begin{align}\label{eq:F12l}
& \mathcal{F}(t_1,t_2)=\frac{i}{\hbar^3}\sum_{a,c}|\mu_{ag}|^2\alpha_{ac}^2\theta(t_1)\theta(t_2)e^{-\gamma_a(t_1+2t_2)}\notag\\
 &\times\left[e^{-i\omega_-t_1}-\frac{2i\delta}{k+2i\delta}e^{-kt_2}\left(e^{-i\omega_-t_1}-e^{-(k+i\omega_+)t_1}\right)\right].
 \end{align}
A vibrational dephasing has been added in Eq. (\ref{eq:F12l}) $e^{-\gamma_a t}$ to $\mathcal{G}_{ac,ac}$ and $e^{-2\gamma_a t}$ added to $\mathcal{G}_{aa,aa}$ which is coming from both bra- and ket- propagators. By changing the variables $t'-\tau$ $t'+\tau$ to $t$ in Eq. (\ref{eq:frji}) and (\ref{eq:frjii}), respectively and exchanging $t'$ and $t$ integration in one of them signals (\ref{eq:frji}) and (\ref{eq:frjii}) become complex conjugate of each other. Expanding the pump pulse $\mathcal{E}_2(t)$ into a frequency domain one can evaluate time integrals. Remaining integrals over $\omega_2$ and $\Delta$ can be evaluated using residue calculus assuming pulse $\mathcal{E}_2$ shorter than the delay $T$.

 The result for FR-SPRS signal then reads
\begin{align}\label{eq:frg}
&S_{FR-SPRS}(\omega,T)=\mathcal{R}\frac{\mathcal{D}^2(\omega)}{\hbar^2}|\mathcal{E}_p|^2\sum_{a,c}\alpha_{ac}^2|\mu_{ag}|^2e^{-2\gamma_aT}\notag\\
&\times\left(\mathcal{E}_2^{*}(\omega-\omega_--i\gamma_a)\mathcal{E}_2(\omega-\omega_-+i\gamma_a)-\frac{2i\delta e^{-kT}}{k+2i\delta}\right.\notag\\
&\left.\left[\mathcal{E}_2^{*}(\omega-\omega_--i\gamma_a)\mathcal{E}_2(\omega-\omega_-+i(\gamma_a+k))\right.\right. \notag\\
&\left.\left.-\mathcal{E}_2^{*}(\omega-\omega_{+}-i(\gamma_a+k))\mathcal{E}_2(\omega-\omega_++i\gamma_a)\right]\right).
\end{align}
After applying the SML to Eq. (\ref{eq:frg}), in the limit of short pulse compare to splitting $\delta$ we obtain Eq. (\ref{eq:SP2}). In the FML Eq. (\ref{eq:frg})  yields
\begin{align}\label{eq:SP2f}
&S_{FR-SPRS}^{(FML)}(\omega,T)\notag\\
&=\frac{\mathcal{D}^2(\omega)}{\hbar^2}|\mathcal{E}_p|^2\sum_{a,c}\alpha_{ac}^2|\mu_{ag}|^2|\mathcal{E}_2(\omega-\omega_{-})|^2e^{-2\gamma_aT}
\end{align}

\subsection{FSRS with TSJ (Low T)}

The time domain signals (\ref{eq:Sisr}) - (\ref{eq:Siisr}) for the impulsive actinic pulse can be recast in Liouville space as follows:

\begin{align}
\tilde{S}_{FSRS}^{(i)}(\omega,T;\Delta)&=\frac{2}{\hbar}\int_{-\infty}^{\infty}dt\int_{-\infty}^td\tau_3|\mathcal{E}_p|^2|\mathcal{E}_2|^2e^{-i\Delta(\tau_3-T)}\notag\\
&\times e^{i(\omega-\omega_2)(t-\tau_3)}\mathcal{F}(t-\tau_3,\tau_3),
\end{align}
\begin{align}
\tilde{S}_{FSRS}^{(ii)}(\omega,T;\Delta)&=\frac{2}{\hbar}\int_{-\infty}^{\infty}dt\int_{-\infty}^td\tau_3|\mathcal{E}_p|^2|\mathcal{E}_2|^2e^{-i\Delta(\tau_3-T)}\notag\\
&\times e^{i(\omega-\omega_2)(t-\tau_3)}\mathcal{F}^{*}(t-\tau_3,\tau_3),
\end{align}


Taking into account Eq. (\ref{eq:FL}), evaluating the time integrals and use residue calculus for evaluating $\Delta$ integral in  Eq. (\ref{eq:SRS}) we obtain for the total FSRS signal
\begin{align}\label{eq:FSRS2g}
&S_{FSRS}(\Omega,T)=-\mathcal{I}\frac{2}{\hbar^4}|\mathcal{E}_p|^2|\mathcal{E}_2|^2\mathcal{E}_3^{*}(\Omega+\omega_2)\sum_{a,c}\alpha_{ac}^2|\mu_{ag}|^2\notag\\
&\times\left[\frac{\mathcal{E}_3(\Omega+\omega_2+2i\gamma_a)}{\Omega-\omega_-+i\gamma_a}-\frac{2i\delta\mathcal{E}_3(\Omega+\omega_2+i(2\gamma_a+k))e^{-kT}}{k+2i\delta}\right.\notag\\
&\left.\times\left(\frac{1}{\Omega-\omega_-+i\gamma_a}-\frac{1}{\Omega-\omega_++i(\gamma_a+k)}\right)\right]e^{-2\gamma_aT}\notag\\
&-[\omega_{\pm}\leftrightarrow-\omega_{\mp}].
\end{align}
where $\Omega=\omega-\omega_2$. In the SML and short pulse limit Eq. (\ref{eq:FSRS2g}) yields Eq. (\ref{eq:FSRS2}). In the FML Eq. (\ref{eq:FSRS2g}) reads
\begin{align}\label{eq:FSRS2f}
&S_{FSRS}^{(FML)}(\Omega,T)=\frac{1}{\hbar^4}|\mathcal{E}_p|^2|\mathcal{E}_2|^2|\mathcal{E}_3(\Omega+\omega_2)|^2\sum_{a,c}\alpha_{ac}^2|\mu_{ag}|^2\notag\\
&\times\left(\frac{2\delta}{k}e^{-kT}\left[\frac{\Omega-\omega_-}{(\Omega-\omega_-)^2+\gamma_a^2}-\frac{\Omega-\omega_+}{(\Omega-\omega_+)^2+(\gamma_a+k)^2}\right]\right.\notag\\
&\left.+\frac{\gamma_a}{(\Omega-\omega_-)^2+\gamma_a^2}\right)e^{-2\gamma_aT}-[\omega_{\pm}\leftrightarrow-\omega_{\mp}].
\end{align}

\subsection{TG-ISRS and TR-ISRS with TSJ (Low T)}

 We recast TG-ISRS in Eqs. (\ref{eq:Stgi}) - (\ref{eq:Stgii}) in Liouville space using the impulsive actinic limit
\begin{align}
\tilde{S}_{TG-ISRS}^{(i)}(\Omega,T)&=\frac{2}{\hbar}\int_{-\infty}^{\infty}dt\int_{-\infty}^td\tau_3|\mathcal{E}_p|^2e^{i\Omega(t-T)}\notag\\
&\times\mathcal{E}_2^{*}(\tau_3-T)\mathcal{E}_1(\tau_3-T)\mathcal{F}(t-\tau_3,\tau_3),
\end{align}
\begin{align}
\tilde{S}_{TG-ISRS}^{(i)}(\Omega,T)&=\frac{2}{\hbar}\int_{-\infty}^{\infty}dt\int_{-\infty}^td\tau_3|\mathcal{E}_p|^2e^{i\Omega(t-T)}\notag\\
&\times\mathcal{E}_2^{*}(\tau_3-T)\mathcal{E}_2(\tau_3-T)\mathcal{F}^{*}(t-\tau_3,\tau_3),
\end{align}
W next expand the fields $\mathcal{E}_1(t)$ and $\mathcal{E}_2(t)$ in frequency domain to evaluate the time integrals and $\Delta$ frequency integral. 

Using Eq. (\ref{eq:FL}) we evaluate the TG-ISRS signal (\ref{eq:TGdef})
\begin{align}\label{eq:TGg}
&S_{TG-ISRS}(\Omega,T)=-\mathcal{I}\frac{2}{\hbar^4}|\mathcal{E}_p|^2\sum_{a,c}\alpha_{ac}^2|\mu_{ag}|^2e^{-2\gamma_aT}\notag\\
&\times\int_{-\infty}^{\infty}\frac{d\omega'}{2\pi}\mathcal{E}_s^{*}(\omega')\mathcal{E}_3(\omega'-\Omega)\int_{-\infty}^{\infty}\frac{d\omega_2}{2\pi}\mathcal{E}_2^{*}(\omega_2)\notag\\
&\times\left[\frac{\mathcal{E}_2(\omega_2+\Omega+2i\gamma_a)}{\Omega-\omega_-+i\gamma_a}-\frac{2i\delta\mathcal{E}_2(\omega_2+\Omega+i(2\gamma_a+k))e^{-kT}}{k+2i\delta}\right.\notag\\
&\left.\times\left(\frac{1}{\Omega-\omega_-+i\gamma_a}-\frac{1}{\Omega-\omega_++i(\gamma_a+k)}\right)\right]-[\omega_{\pm}\leftrightarrow-\omega_{\mp}]
\end{align}
In the SML and short pulse limit Eq. (\ref{eq:TGg}) yields Eq. (\ref{eq:TG2}). Similarly we obtain the TSJ expression for the TR-ISRS signal by replacing $\mathcal{E}_s\to\mathcal{E}_3$ and $\mathcal{E}_1\to\mathcal{E}_2$ in Eq. (\ref{eq:TGg}). In the FML Eq. (\ref{eq:TGg}) yields
\begin{align}\label{eq:TG2f}
&S_{TG-ISRS}^{(FML)}(\Omega,T)=\frac{1}{\hbar^4}|\mathcal{E}_p|^2\sum_{a,c}\alpha_{ac}^2|\mu_{ag}|^2\notag\\
&\times\int_{-\infty}^{\infty}\frac{d\omega'}{2\pi}\mathcal{E}_s(\omega')\mathcal{E}_3(\omega'-\Omega)\int_{-\infty}^{\infty}\frac{d\omega_2}{2\pi}\mathcal{E}_2(\omega_2)\mathcal{E}_1(\omega_2+\Omega)\notag\\
&\times\left(\frac{2\delta}{k}e^{-kT}\left[\frac{\Omega-\omega_-}{(\Omega-\omega_-)^2+\gamma_a^2}-\frac{\Omega-\omega_+}{(\Omega-\omega_+)^2+(\gamma_a+k)^2}\right]\right.\notag\\
&\left.+\frac{\gamma_a}{(\Omega-\omega_-)^2+\gamma_a^2}\right)e^{-2\gamma_aT}-[\omega_{\pm}\leftrightarrow-\omega_{\mp}].
\end{align}

\section{High temperature limit}\label{sec:TSJh}

\subsection{TSJ model in high temperature limit}

In the high temperature limit $k_BT\gg\epsilon_u-\epsilon_d$ Eqs. (\ref{eq:Gaa}) - (\ref{eq:Gac}) yield
\begin{align}
\mathcal{G}_{aa,aa}(t)=(-i/\hbar)\theta(t)e^{-kt}
 \begin{pmatrix}
  \cosh(k t)& \sinh(k t) \\
  \sinh(k t) & \cosh(k t) \\
 \end{pmatrix},
 \end{align}
\begin{align}
&\mathcal{G}_{ac,ac}(t)=(-i/\hbar)\theta(t)e^{-(k+i\omega_{ac})t}\notag\\
&\times
 \begin{pmatrix}
  \cosh(\eta t)-i\frac{\delta}{\eta}\sinh(\eta t)& \frac{k}{\eta}\sinh(\eta t) \\
  \frac{k}{\eta}\sinh(\eta t) & \cosh(\eta t)+i\frac{\delta}{\eta}\sinh(\eta t) \\
 \end{pmatrix},
 \end{align}
where $k=k_u=k_d$ and $\eta=\sqrt{k^2-\delta^2}$. The linear absorption (\ref{eq:Sl}) is then given by
\begin{align}\label{eq:Sl1}
S_l(\omega)&=\frac{2}{\hbar^2\eta}(k^2-\eta^2)|\mathcal{E}(\omega)|^2|\mu_{ac}|^2\notag\\
&\times\left[\frac{1}{(\omega-\omega_{ac})^2+(k-\eta)^2}-\frac{1}{(\omega-\omega_{ac})^2+(k+\eta)^2}\right].
\end{align}
In the SML the signal (\ref{eq:Sl1}) gives two distinct narrow peaks with central frequencies $\omega=\omega_{ac}\pm\delta$ with width $\delta$
\begin{align}\label{eq:Slsml2}
S_l^{(SML)}(\omega)&=\frac{2}{\hbar^2}k|\mathcal{E}(\omega)|^2|\mu_{ac}|^2\notag\\
&\times\left[\frac{1}{(\omega-\omega_{ac}-\delta)^2+k^2}+\frac{1}{(\omega-\omega_{ac}+\delta)^2+k^2}\right].
\end{align}
In the opposite FML Eq. (\ref{eq:Sl1}) gives a single peak at $\omega=\omega_{ac}$ with width $\delta^2/2k$
\begin{align}
S_l^{(FML)}(\omega)&=\frac{2}{\hbar^2}\frac{\delta^2}{k}\frac{|\mathcal{E}(\omega)|^2|\mu_{ac}|^2}{(\omega-\omega_{ac})^2+(\delta^2/2k)^2}.
\end{align}
Therefore unlike the low temperature behavior described by (\ref{eq:Slsml}) the absorption spectrum (\ref{eq:Slsml2}) consists of two, rather than one peaks that governs the equilibrium population of both $u$ and $d$ states.

We now turn to the Raman signals. For the general relation between jump rate $k$ and splitting $\delta$ Eq. (\ref{eq:FL}) can be recast as
 \begin{align}\label{eq:F12}
 \mathcal{F}(t_1,t_2)&=\frac{i}{\hbar^3}\sum_{a,c}\alpha_{ac}^2|\mu_{ag}|^2\theta(t_1)\theta(t_2)\notag\\
 &\times\sum_{\mu,\nu=\pm}A_{\mu\nu}e^{-[\omega_{ac}+\gamma_\mu+\gamma_a]t_1-(\Gamma_\nu+2\gamma_a) t_2},
 \end{align}
where $A_{\pm+}=\frac{1}{2}\left(1\pm\frac{k}{\eta}\right)$, $A_{\pm-}=\mp i\frac{\delta}{2\eta}$, and $\gamma_{\pm}=k\mp\eta$, $\Gamma_-=0$, and $\Gamma_+=2k$.

\subsection{FR-SPRS with TSJ (High T)}

In the high temperature limit the matter $\mathcal{F}(t_1,t_2)$  is given by Eq. (\ref{eq:F12}).  The result for FR-SPRS signal then reads
\begin{align}\label{eq:frgh}
S_{FR-SPRS}&(\omega,T)=\mathcal{R}\frac{\mathcal{D}^2(\omega)}{\hbar^2}|\mathcal{E}_p|^2\sum_{a,c}\alpha_{ac}^2|\mu_{ag}|^2e^{-2\gamma_aT}\notag\\
\times&\left(\left[1+\frac{k-i\delta e^{-2kT}}{\eta}\right]|\mathcal{E}_2(\omega-\omega_{ac}-i\eta)|^2\right.\notag\\
&\left.+\left[1-\frac{k-i\delta e^{-2kT}}{\eta}\right]|\mathcal{E}_2(\omega-\omega_{ac}+i\eta)|^2\right),
\end{align}
where we assumed that the pulse is short compare to $\gamma_a^{-1}$ and $k^{-1}$. In the SML  Eq. (\ref{eq:frgh}), yields 
\begin{align}\label{eq:SP2h}
&S_{FR-SPRS}^{(SML)}(\omega,T)=\frac{\mathcal{D}^2(\omega)}{\hbar^2}|\mathcal{E}_p|^2\sum_{a,c}\alpha_{ac}^2|\mu_{ag}|^2e^{-2\gamma_aT}\notag\\
&\times \left(|\mathcal{E}_2(\omega-\omega_{-})|^2[1-e^{-2kT}]+|\mathcal{E}_2(\omega-\omega_+)|^2[1+e^{-2kT}]\right).
\end{align}
In the FML Eq. (\ref{eq:frgh})  reads
\begin{align}\label{eq:SP2f}
&S_{FR-SPRS}^{(FML)}(\omega,T)\notag\\
&=\frac{\mathcal{D}^2(\omega)}{\hbar^2}|\mathcal{E}_p|^2\sum_{a,c}\alpha_{ac}^2|\mu_{ag}|^2|\mathcal{E}_2(\omega-\omega_{ac})|^2e^{-2\gamma_aT}.
\end{align}

\subsection{FSRS with TSJ (High T)}

Using Eq. (\ref{eq:F12}) for a short pulse compare to $k^{-1}$ and $\gamma_a^{-1}$ we obtain for the total FSRS signal
\begin{align}\label{eq:FSRS2gh}
S_{FSRS}&(\Omega,T)=-\mathcal{I}\frac{1}{\hbar^4}|\mathcal{E}_p|^2|\mathcal{E}_2|^2|\mathcal{E}_3(\Omega+\omega_2)|^2\sum_{a,c}\alpha_{ac}^2|\mu_{ag}|^2\notag\\
\times&\left[\left(1+\frac{k-i\delta e^{-2kT}}{\eta}\right)\frac{1}{\Omega-\omega_{ac}+i(\gamma-\eta)}\right.\notag\\
&\left.\left(1-\frac{k-i\delta e^{-2kT}}{\eta}\right)\frac{1}{\Omega-\omega_{ac}+i(\gamma+\eta)}\right]e^{-2\gamma_aT}\notag\\
&-[\omega_{ac}\leftrightarrow-\omega_{ac}]
\end{align}
In the SML and short pulse limit Eq. (\ref{eq:FSRS2gh}) yields 
\begin{align}\label{eq:FSRS2h}
&S_{FSRS}^{(SML)}(\Omega,T)=\frac{1}{\hbar^4}|\mathcal{E}_p|^2|\mathcal{E}_2|^2|\mathcal{E}_3(\Omega+\omega_2)|^2\sum_{a,c}\alpha_{ac}^2|\mu_{ag}|^2\notag\\
&\times \left(\gamma(1-e^{-2kT})\left[\frac{1}{(\Omega-\omega_-)^2+\gamma^2}+\frac{1}{(\Omega-\omega_+)^2+\gamma^2}\right]\right.\notag\\
&\left.+\frac{k}{\delta}\left[\frac{\Omega-\omega_-}{(\Omega-\omega_-)^2+\gamma^2}-\frac{\Omega-\omega_+}{(\Omega-\omega_+)^2+\gamma^2}\right]\right)e^{-2\gamma_aT}\notag\\
&-[\omega_{\pm}\leftrightarrow-\omega_{\mp}],
\end{align}
where $\gamma=\gamma_a+k$. In the FML Eq. (\ref{eq:FSRS2gh}) reads
\begin{align}\label{eq:FSRS2fh}
&S_{FSRS}^{(FML)}(\Omega,T)=\frac{1}{\hbar^4}|\mathcal{E}_p|^2|\mathcal{E}_2|^2|\mathcal{E}_3(\Omega+\omega_2)|^2\sum_{a,c}\alpha_{ac}^2|\mu_{ag}|^2\notag\\
&\times\frac{2\left(\frac{\delta^2}{2k}+\gamma_a\right)+\frac{\delta}{k}e^{-2kT}(\Omega-\omega_{ac})}{(\Omega-\omega_{ac})^2+\left(\frac{\delta^2}{2k}+\gamma_a\right)^2}-[\omega_{ac}\leftrightarrow-\omega_{ac}],
\end{align}

\subsection{TG-ISRS and TR-ISRS with TSJ (High T)}

Using Eq. (\ref{eq:F12}) we evaluate the TG-ISRS signal (\ref{eq:TGdef})
\begin{align}\label{eq:TGgh}
&S_{TG-ISRS}(\Omega,T)=-\mathcal{I}\frac{2}{\hbar^4}|\mathcal{E}_p|^2\sum_{a,c}\alpha_{ac}^2|\mu_{ag}|^2e^{-2\gamma_aT}\notag\\
&\times\int_{-\infty}^{\infty}\frac{d\omega'}{2\pi}\mathcal{E}_s^{*}(\omega')\mathcal{E}_3(\omega'-\Omega)\int_{-\infty}^{\infty}\frac{d\omega_2}{2\pi}\mathcal{E}_2^{*}(\omega_2)\mathcal{E}_2(\omega_2+\Omega)\notag\\
&\times\left[\left(1+\frac{k-i\delta e^{-2kT}}{\eta}\right)\frac{1}{\Omega-\omega_{ac}+i(\gamma-\eta)}\right.\notag\\
&\left.\left(1-\frac{k-i\delta e^{-2kT}}{\eta}\right)\frac{1}{\Omega-\omega_{ac}+i(\gamma+\eta)}\right]e^{-2\gamma_aT}\notag\\
&-[\omega_{ac}\leftrightarrow-\omega_{ac}]
\end{align}
In the SML and short pulse limit Eq. (\ref{eq:TGgh}) yields 
\begin{align}\label{eq:TG2h}
&S_{TG-ISRS}^{(SML)}(\Omega,T)=\frac{1}{\hbar^4}|\mathcal{E}_p|^2\sum_{a,c}\alpha_{ac}^2|\mu_{ag}|^2\notag\\
&\times\int_{-\infty}^{\infty}\frac{d\omega'}{2\pi}\mathcal{E}_s(\omega')\mathcal{E}_3(\omega'-\Omega)\int_{-\infty}^{\infty}\frac{d\omega_2}{2\pi}\mathcal{E}_2(\omega_2)\mathcal{E}_1(\omega_2+\Omega)\notag\\
&\times \left(\gamma(1-e^{-2kT})\left[\frac{1}{(\Omega-\omega_-)^2+\gamma^2}+\frac{1}{(\Omega-\omega_+)^2+\gamma^2}\right]\right.\notag\\
&\left.+\frac{k}{\delta}\left[\frac{\Omega-\omega_-}{(\Omega-\omega_-)^2+\gamma^2}-\frac{\Omega-\omega_+}{(\Omega-\omega_+)^2+\gamma^2}\right]\right)e^{-2\gamma_aT}\notag\\
&-[\omega_{\pm}\leftrightarrow-\omega_{\mp}].
\end{align}
 In the FML Eq. (\ref{eq:TGgh}) yields
\begin{align}\label{eq:TG2f}
&S_{TG-ISRS}^{(FML)}(\Omega,T)=\frac{1}{\hbar^4}|\mathcal{E}_p|^2\sum_{a,c}\alpha_{ac}^2|\mu_{ag}|^2\notag\\
&\times\int_{-\infty}^{\infty}\frac{d\omega'}{2\pi}\mathcal{E}_s(\omega')\mathcal{E}_3(\omega'-\Omega)\int_{-\infty}^{\infty}\frac{d\omega_2}{2\pi}\mathcal{E}_2(\omega_2)\mathcal{E}_1(\omega_2+\Omega)\notag\\
&\times\frac{2\left(\frac{\delta^2}{2k}+\gamma_a\right)+\frac{\delta}{k}e^{-2kT}(\Omega-\omega_{ac})}{(\Omega-\omega_{ac})^2+\left(\frac{\delta^2}{2k}+\gamma_a\right)^2}-[\omega_{ac}\leftrightarrow-\omega_{ac}].
\end{align}
The corresponding TR-ISRS are given by Eqs. (\ref{eq:TGgh}) - (\ref{eq:TG2f}) by  simply replacing $\mathcal{E}_s\to\mathcal{E}_3$ and $\mathcal{E}_1\to\mathcal{E}_2$.

%


\begin{thebibliography}{10}%
\makeatletter
\providecommand \@ifxundefined [1]{%
 \ifx #1\undefined \expandafter \@firstoftwo
 \else \expandafter \@secondoftwo
\fi
}%
\providecommand \@ifnum [1]{%
 \ifnum #1\expandafter \@firstoftwo
 \else \expandafter \@secondoftwo
\fi
}%
\providecommand \enquote [1]{``#1''}%
\providecommand \bibnamefont  [1]{#1}%
\providecommand \bibfnamefont [1]{#1}%
\providecommand \citenamefont [1]{#1}%
\providecommand\href[0]{\@sanitize\@href}%
\providecommand\@href[1]{\endgroup\@@startlink{#1}\endgroup\@@href}%
\providecommand\@@href[1]{#1\@@endlink}%
\providecommand \@sanitize [0]{\begingroup\catcode`\&12\catcode`\#12\relax}%
\@ifxundefined \pdfoutput {\@firstoftwo}{%
 \@ifnum{\z@=\pdfoutput}{\@firstoftwo}{\@secondoftwo}%
}{%
 \providecommand\@@startlink[1]{\leavevmode}%
 \providecommand\@@endlink[0]{}%
}{%
 \providecommand\@@startlink[1]{%
  \leavevmode
  \pdfstartlink
   attr{/Border[0 0 1 ]/H/I/C[0 1 1]}%
   user{/Subtype/Link/A<</Type/Action/S/URI/URI(#1)>>}%
  \relax
 }%
 \providecommand\@@endlink[0]{\pdfendlink}%
}%
\providecommand \url  [0]{\begingroup\@sanitize \@url }%
\providecommand \@url [1]{\endgroup\@href {#1}{\urlprefix}}%
\providecommand \urlprefix [0]{URL }%
\providecommand \Eprint[0]{\href }%
\@ifxundefined \urlstyle {%
  \providecommand \doi [1]{doi:\discretionary{}{}{}#1}%
}{%
  \providecommand \doi [0]{doi:\discretionary{}{}{}\begingroup
  \urlstyle{rm}\Url }%
}%
\providecommand \doibase [0]{http://dx.doi.org/}%
\providecommand \Doi[1]{\href{\doibase#1}}%
\providecommand \bibAnnote [3]{%
  \BibitemShut{#1}%
  \begin{quotation}\noindent
    \textsc{Key:}\ #2\\\textsc{Annotation:}\ #3%
  \end{quotation}%
}%
\providecommand \bibAnnoteFile [2]{%
  \IfFileExists{#2}{\bibAnnote {#1} {#2} {\input{#2}}}{}%
}%
\providecommand \typeout [0]{\immediate \write \m@ne }%
\providecommand \selectlanguage [0]{\@gobble}%
\providecommand \bibinfo [0]{\@secondoftwo}%
\providecommand \bibfield [0]{\@secondoftwo}%
\providecommand \translation [1]{[#1]}%
\providecommand \BibitemOpen[0]{}%
\providecommand \bibitemStop [0]{}%
\providecommand \bibitemNoStop [0]{.\EOS\space}%
\providecommand \EOS [0]{\spacefactor3000\relax}%
\providecommand \BibitemShut [1]{\csname bibitem#1\endcsname}%
\bibitem{Miz97}%
  \BibitemOpen
  \bibfield{author}{%
  \bibinfo {author} {\bibfnamefont{Y.}~\bibnamefont{Mizutani}}\ and\ \bibinfo
  {author} {\bibfnamefont{T.}~\bibnamefont{Kitagawa}},\ }%
  \bibfield{journal}{%
  \bibinfo {journal} {Science}\ }%
  \textbf{\bibinfo {volume} {278}},\ \bibinfo {pages} {443} (\bibinfo {year}
  {1997})%
  \bibAnnoteFile{NoStop}{Miz97}%
\bibitem{McCamant:JPCA:2003}%
  \BibitemOpen
  \bibfield{author}{%
  \bibinfo {author} {\bibfnamefont{D.~W.}\ \bibnamefont{McCamant}}, \bibinfo
  {author} {\bibfnamefont{P.}~\bibnamefont{Kukura}},\ and\ \bibinfo {author}
  {\bibfnamefont{R.~A.}\ \bibnamefont{Mathies}},\ }%
  \bibfield{journal}{%
  \bibinfo {journal} {The Journal of Physical Chemistry A}\ }%
  \textbf{\bibinfo {volume} {107}},\ \bibinfo {pages} {8208} (\bibinfo {year}
  {2003})%
  \bibAnnoteFile{NoStop}{McCamant:JPCA:2003}%
\bibitem{Lee:JCP:2004}%
  \BibitemOpen
  \bibfield{author}{%
  \bibinfo {author} {\bibfnamefont{S.-Y.}\ \bibnamefont{Lee}}, \bibinfo
  {author} {\bibfnamefont{D.}~\bibnamefont{Zhang}}, \bibinfo {author}
  {\bibfnamefont{D.~W.}\ \bibnamefont{McCamant}}, \bibinfo {author}
  {\bibfnamefont{P.}~\bibnamefont{Kukura}},\ and\ \bibinfo {author}
  {\bibfnamefont{R.~A.}\ \bibnamefont{Mathies}},\ }%
  \bibfield{journal}{%
  \bibinfo {journal} {The Journal of Chemical Physics}\ }%
  \textbf{\bibinfo {volume} {121}},\ \bibinfo {pages} {3632} (\bibinfo {year}
  {2004})%
  \bibAnnoteFile{NoStop}{Lee:JCP:2004}%
\bibitem{Fang:Nature:2009}%
  \BibitemOpen
  \bibfield{author}{%
  \bibinfo {author} {\bibfnamefont{C.}~\bibnamefont{Fang}}, \bibinfo {author}
  {\bibfnamefont{R.~R.}\ \bibnamefont{Frontiera}}, \bibinfo {author}
  {\bibfnamefont{R.}~\bibnamefont{Tran}},\ and\ \bibinfo {author}
  {\bibfnamefont{R.~a.}\ \bibnamefont{Mathies}},\ }%
  \bibfield{journal}{%
  \bibinfo {journal} {Nature}\ }%
  \textbf{\bibinfo {volume} {462}},\ \bibinfo {pages} {200} (\bibinfo {year}
  {2009})%
  \bibAnnoteFile{NoStop}{Fang:Nature:2009}%
\bibitem{Kukura:Science:2005}%
  \BibitemOpen
  \bibfield{author}{%
  \bibinfo {author} {\bibfnamefont{P.}~\bibnamefont{Kukura}}, \bibinfo {author}
  {\bibfnamefont{D.~W.}\ \bibnamefont{McCamant}}, \bibinfo {author}
  {\bibfnamefont{S.}~\bibnamefont{Yoon}}, \bibinfo {author}
  {\bibfnamefont{D.~B.}\ \bibnamefont{Wandschneider}},\ and\ \bibinfo {author}
  {\bibfnamefont{R.~A.}\ \bibnamefont{Mathies}},\ }%
  \bibfield{journal}{%
  \bibinfo {journal} {Science}\ }%
  \textbf{\bibinfo {volume} {310}},\ \bibinfo {pages} {1006} (\bibinfo {year}
  {2005})%
  \bibAnnoteFile{NoStop}{Kukura:Science:2005}%
\bibitem{Kukura:AnnurevPhysChem:2007}%
  \BibitemOpen
  \bibfield{author}{%
  \bibinfo {author} {\bibfnamefont{P.}~\bibnamefont{Kukura}}, \bibinfo {author}
  {\bibfnamefont{D.~W.}\ \bibnamefont{McCamant}},\ and\ \bibinfo {author}
  {\bibfnamefont{R.~A.}\ \bibnamefont{Mathies}},\ }%
  \bibfield{journal}{%
  \bibinfo {journal} {Annual Review of Physical Chemistry}\ }%
  \textbf{\bibinfo {volume} {58}},\ \bibinfo {pages} {461} (\bibinfo {year}
  {2007})%
  \bibAnnoteFile{NoStop}{Kukura:AnnurevPhysChem:2007}%
\bibitem{Takeuchi:Science:2008}%
  \BibitemOpen
  \bibfield{author}{%
  \bibinfo {author} {\bibfnamefont{S.}~\bibnamefont{Takeuchi}}, \bibinfo
  {author} {\bibfnamefont{S.}~\bibnamefont{Ruhman}}, \bibinfo {author}
  {\bibfnamefont{T.}~\bibnamefont{Tsuneda}}, \bibinfo {author}
  {\bibfnamefont{M.}~\bibnamefont{Chiba}}, \bibinfo {author}
  {\bibfnamefont{T.}~\bibnamefont{Taketsugu}},\ and\ \bibinfo {author}
  {\bibfnamefont{T.}~\bibnamefont{Tahara}},\ }%
  \bibfield{journal}{%
  \bibinfo {journal} {Science}\ }%
  \textbf{\bibinfo {volume} {322}},\ \bibinfo {pages} {1073} (\bibinfo {year}
  {2008})%
  \bibAnnoteFile{NoStop}{Takeuchi:Science:2008}%
\bibitem{Kur12}%
  \BibitemOpen
  \bibfield{author}{%
  \bibinfo {author} {\bibfnamefont{H.}~\bibnamefont{Kuramochi}}, \bibinfo
  {author} {\bibfnamefont{S.}~\bibnamefont{Takeuchi}},\ and\ \bibinfo {author}
  {\bibfnamefont{T.}~\bibnamefont{Tahara}},\ }%
  \bibfield{journal}{%
  \bibinfo {journal} {The Journal of Physical Chemistry Letters}\ }%
  \textbf{\bibinfo {volume} {3}},\ \bibinfo {pages} {2025} (\bibinfo {year}
  {2012})%
  \bibAnnoteFile{NoStop}{Kur12}%
\bibitem{Zan09}%
  \BibitemOpen
  \bibfield{author}{%
  \bibinfo {author} {\bibfnamefont{S.-H.}\ \bibnamefont{Shim}}\ and\ \bibinfo
  {author} {\bibfnamefont{M.~T.}\ \bibnamefont{Zanni}},\ }%
  \bibfield{journal}{%
  \bibinfo {journal} {Phys. Chem. Chem. Phys.}\ }%
  \textbf{\bibinfo {volume} {11}},\ \bibinfo {pages} {748} (\bibinfo {year}
  {2009})%
  \bibAnnoteFile{NoStop}{Zan09}%
\bibitem{biggs12}%
  \BibitemOpen
  \bibfield{author}{%
  \bibinfo {author} {\bibfnamefont{J.~D.}\ \bibnamefont{Biggs}}, \bibinfo
  {author} {\bibfnamefont{Y.}~\bibnamefont{Zhang}}, \bibinfo {author}
  {\bibfnamefont{D.}~\bibnamefont{Healion}},\ and\ \bibinfo {author}
  {\bibfnamefont{S.}~\bibnamefont{Mukamel}},\ }%
  \bibfield{journal}{%
  \bibinfo {journal} {The Journal of Chemical Physics}\ }%
  \textbf{\bibinfo {volume} {136}},\ \bibinfo {pages} {174117} (\bibinfo {year}
  {2012})%
  \bibAnnoteFile{NoStop}{biggs12}%
\bibitem{biggs13}%
  \BibitemOpen
  \bibfield{author}{%
  \bibinfo {author} {\bibfnamefont{S.}~\bibnamefont{Mukamel}}, \bibinfo
  {author} {\bibfnamefont{D.}~\bibnamefont{Healion}}, \bibinfo {author}
  {\bibfnamefont{Y.}~\bibnamefont{Zhang}},\ and\ \bibinfo {author}
  {\bibfnamefont{J.~D.}\ \bibnamefont{Biggs}},\ }%
  \bibfield{journal}{%
  \bibinfo {journal} {Annu. Rev. Phys. Chem.}\ }%
  \textbf{\bibinfo {volume} {64}},\ \bibinfo {pages} {101} (\bibinfo {year}
  {2013})%
  \bibAnnoteFile{NoStop}{biggs13}%
\bibitem{Ham94}%
  \BibitemOpen
  \bibfield{author}{%
  \bibinfo {author} {\bibfnamefont{H.}~\bibnamefont{Hamaguchi}}\ and\ \bibinfo
  {author} {\bibfnamefont{T.~L.}\ \bibnamefont{Gustafson}},\ }%
  \bibfield{journal}{%
  \bibinfo {journal} {Annual Review of Physical Chemistry}\ }%
  \textbf{\bibinfo {volume} {45}},\ \bibinfo {pages} {593} (\bibinfo {year}
  {1994})%
  \bibAnnoteFile{NoStop}{Ham94}%
\bibitem{Tat00}%
  \BibitemOpen
  \bibfield{author}{%
  \bibinfo {author} {\bibfnamefont{T.}~\bibnamefont{Fujino}}\ and\ \bibinfo
  {author} {\bibfnamefont{T.}~\bibnamefont{Tahara}},\ }%
  \bibfield{journal}{%
  \bibinfo {journal} {The Journal of Physical Chemistry A}\ }%
  \textbf{\bibinfo {volume} {104}},\ \bibinfo {pages} {4203} (\bibinfo {year}
  {2000})%
  \bibAnnoteFile{NoStop}{Tat00}%
\bibitem{Kim01}%
  \BibitemOpen
  \bibfield{author}{%
  \bibinfo {author} {\bibfnamefont{J.~E.}\ \bibnamefont{Kim}}, \bibinfo
  {author} {\bibfnamefont{D.~W.}\ \bibnamefont{McCamant}}, \bibinfo {author}
  {\bibfnamefont{L.}~\bibnamefont{Zhu}},\ and\ \bibinfo {author}
  {\bibfnamefont{R.~A.}\ \bibnamefont{Mathies}},\ }%
  \bibfield{journal}{%
  \bibinfo {journal} {The Journal of Physical Chemistry B}\ }%
  \textbf{\bibinfo {volume} {105}},\ \bibinfo {pages} {1240} (\bibinfo {year}
  {2001})%
  \bibAnnoteFile{NoStop}{Kim01}%
\bibitem{Fuj03}%
  \BibitemOpen
  \bibfield{author}{%
  \bibinfo {author} {\bibfnamefont{S.}~\bibnamefont{Fujiyoshi}}, \bibinfo
  {author} {\bibfnamefont{S.}~\bibnamefont{Takeuchi}},\ and\ \bibinfo {author}
  {\bibfnamefont{T.}~\bibnamefont{Tahara}},\ }%
  \bibfield{journal}{%
  \bibinfo {journal} {The Journal of Physical Chemistry A}\ }%
  \textbf{\bibinfo {volume} {107}},\ \bibinfo {pages} {494} (\bibinfo {year}
  {2003})%
  \bibAnnoteFile{NoStop}{Fuj03}%
\bibitem{Kra13}%
  \BibitemOpen
  \bibfield{author}{%
  \bibinfo {author} {\bibfnamefont{J.~P.}\ \bibnamefont{Kraack}}, \bibinfo
  {author} {\bibfnamefont{A.}~\bibnamefont{Wand}}, \bibinfo {author}
  {\bibfnamefont{T.}~\bibnamefont{Buckup}}, \bibinfo {author}
  {\bibfnamefont{M.}~\bibnamefont{Motzkus}},\ and\ \bibinfo {author}
  {\bibfnamefont{S.}~\bibnamefont{Ruhman}},\ }%
  \bibfield{journal}{%
  \Doi{10.1039/C3CP50871D}{\bibinfo {journal} {Phys. Chem. Chem. Phys.}},\ }%
   (\bibinfo {year} {2013}),\ \url{http://dx.doi.org/10.1039/C3CP50871D}%
  \bibAnnoteFile{NoStop}{Kra13}%
\bibitem{Kno91}%
  \BibitemOpen
  \bibfield{author}{%
  \bibinfo {author} {\bibfnamefont{J.}~\bibnamefont{Knoester}}\ and\ \bibinfo
  {author} {\bibfnamefont{S.}~\bibnamefont{Mukamel}},\ }%
  \bibfield{journal}{%
  \bibinfo {journal} {Physics Reports}\ }%
  \textbf{\bibinfo {volume} {205}},\ \bibinfo {pages} {1 } (\bibinfo {year}
  {1991})%
  \bibAnnoteFile{NoStop}{Kno91}%
\bibitem{Hog96}%
  \BibitemOpen
  \bibfield{author}{%
  \bibinfo {author} {\bibfnamefont{C.}~\bibnamefont{Hogemann}}, \bibinfo
  {author} {\bibfnamefont{M.}~\bibnamefont{Pauchard}},\ and\ \bibinfo {author}
  {\bibfnamefont{E.}~\bibnamefont{Vauthey}},\ }%
  \bibfield{journal}{%
  \bibinfo {journal} {Review of Scientific Instruments}\ }%
  \textbf{\bibinfo {volume} {67}},\ \bibinfo {pages} {3449} (\bibinfo {year}
  {1996})%
  \bibAnnoteFile{NoStop}{Hog96}%
\bibitem{Goo98}%
  \BibitemOpen
  \bibfield{author}{%
  \bibinfo {author} {\bibfnamefont{G.~D.}\ \bibnamefont{Goodno}}, \bibinfo
  {author} {\bibfnamefont{G.}~\bibnamefont{Dadusc}},\ and\ \bibinfo {author}
  {\bibfnamefont{R.~J.~D.}\ \bibnamefont{Miller}},\ }%
  \bibfield{journal}{%
  \bibinfo {journal} {J. Opt. Soc. Am. B}\ }%
  \textbf{\bibinfo {volume} {15}},\ \bibinfo {pages} {1791} (\bibinfo {year}
  {1998})%
  \bibAnnoteFile{NoStop}{Goo98}%
\bibitem{Xu01}%
  \BibitemOpen
  \bibfield{author}{%
  \bibinfo {author} {\bibfnamefont{Q.-H.}\ \bibnamefont{Xu}}, \bibinfo {author}
  {\bibfnamefont{Y.-Z.}\ \bibnamefont{Ma}},\ and\ \bibinfo {author}
  {\bibfnamefont{G.~R.}\ \bibnamefont{Fleming}},\ }%
  \bibfield{journal}{%
  \bibinfo {journal} {Chemical Physics Letters}\ }%
  \textbf{\bibinfo {volume} {338}},\ \bibinfo {pages} {254 } (\bibinfo {year}
  {2001})%
  \bibAnnoteFile{NoStop}{Xu01}%
\bibitem{Rhi10}%
  \BibitemOpen
  \bibfield{author}{%
  \bibinfo {author} {\bibfnamefont{J.~M.}\ \bibnamefont{Rhinehart}}, \bibinfo
  {author} {\bibfnamefont{R.~D.}\ \bibnamefont{Mehlenbacher}},\ and\ \bibinfo
  {author} {\bibfnamefont{D.}~\bibnamefont{McCamant}},\ }%
  \bibfield{journal}{%
  \bibinfo {journal} {The Journal of Physical Chemistry B}\ }%
  \textbf{\bibinfo {volume} {114}},\ \bibinfo {pages} {14646} (\bibinfo {year}
  {2010})%
  \bibAnnoteFile{NoStop}{Rhi10}%
\bibitem{McC11}%
  \BibitemOpen
  \bibfield{author}{%
  \bibinfo {author} {\bibfnamefont{D.~W.}\ \bibnamefont{McCamant}},\ }%
  \bibfield{journal}{%
  \bibinfo {journal} {The Journal of Physical Chemistry B}\ }%
  \textbf{\bibinfo {volume} {115}},\ \bibinfo {pages} {9299} (\bibinfo {year}
  {2011})%
  \bibAnnoteFile{NoStop}{McC11}%
\bibitem{Kov13}%
  \BibitemOpen
  \bibfield{author}{%
  \bibinfo {author} {\bibfnamefont{S.~A.}\ \bibnamefont{Kovalenko}}, \bibinfo
  {author} {\bibfnamefont{A.~L.}\ \bibnamefont{Dobryakov}}, \bibinfo {author}
  {\bibfnamefont{E.}~\bibnamefont{Pollak}},\ and\ \bibinfo {author}
  {\bibfnamefont{N.~P.}\ \bibnamefont{Ernsting}},\ }%
  \bibfield{journal}{%
  \bibinfo {journal} {The Journal of Chemical Physics}\ }%
  \textbf{\bibinfo {volume} {139}},\ \bibinfo {eid} {011101} (\bibinfo {year}
  {2013})%
  \bibAnnoteFile{NoStop}{Kov13}%
\bibitem{Pon13}%
  \BibitemOpen
  \bibfield{author}{%
  \bibinfo {author} {\bibfnamefont{E.}~\bibnamefont{Pontecorvo}}, \bibinfo
  {author} {\bibfnamefont{C.}~\bibnamefont{Ferrante}}, \bibinfo {author}
  {\bibfnamefont{C.~G.}\ \bibnamefont{Elles}},\ and\ \bibinfo {author}
  {\bibfnamefont{T.}~\bibnamefont{Scopigno}},\ }%
  \bibfield{journal}{%
  \bibinfo {journal} {Opt. Express}\ }%
  \textbf{\bibinfo {volume} {21}},\ \bibinfo {pages} {6866} (\bibinfo {year}
  {2013})%
  \bibAnnoteFile{NoStop}{Pon13}%
\bibitem{Dor13}%
  \BibitemOpen
  \bibfield{author}{%
  \bibinfo {author} {\bibfnamefont{K.~E.}\ \bibnamefont{Dorfman}}, \bibinfo
  {author} {\bibfnamefont{B.~P.}\ \bibnamefont{Fingerhut}},\ and\ \bibinfo
  {author} {\bibfnamefont{S.}~\bibnamefont{Mukamel}},\ }%
  \bibfield{journal}{%
  \bibinfo {journal} {Phys. Chem. Chem. Phys.}\ }%
  \textbf{\bibinfo {volume} {15}},\ \bibinfo {pages} {12348} (\bibinfo {year}
  {2013})%
  \bibAnnoteFile{NoStop}{Dor13}%
\bibitem{Rah10}%
  \BibitemOpen
  \bibfield{author}{%
  \bibinfo {author} {\bibfnamefont{S.}~\bibnamefont{Rahav}}\ and\ \bibinfo
  {author} {\bibfnamefont{S.}~\bibnamefont{Mukamel}},\ }%
  \bibfield{journal}{%
  \bibinfo {journal} {Sec.4 in Adv. At. Mol., Opt. Phys.}\ }%
  \textbf{\bibinfo {volume} {59}},\ \bibinfo {pages} {223} (\bibinfo {year}
  {2010})%
  \bibAnnoteFile{NoStop}{Rah10}%
\bibitem{Fin13}%
  \BibitemOpen
  \bibfield{author}{%
  \bibinfo {author} {\bibfnamefont{B.~P.}\ \bibnamefont{Fingerhut}}, \bibinfo
  {author} {\bibfnamefont{K.~E.}\ \bibnamefont{Dorfman}},\ and\ \bibinfo
  {author} {\bibfnamefont{S.}~\bibnamefont{Mukamel}},\ }%
  \bibfield{journal}{%
  \bibinfo {journal} {The Journal of Physical Chemistry Letters}\ }%
  \textbf{\bibinfo {volume} {4}},\ \bibinfo {pages} {1933} (\bibinfo {year}
  {2013})%
  \bibAnnoteFile{NoStop}{Fin13}%
\bibitem{Dha94}%
  \BibitemOpen
  \bibfield{author}{%
  \bibinfo {author} {\bibfnamefont{L.}~\bibnamefont{Dhar}}, \bibinfo {author}
  {\bibfnamefont{J.~A.}\ \bibnamefont{Rogers}},\ and\ \bibinfo {author}
  {\bibfnamefont{K.~A.}\ \bibnamefont{Nelson}},\ }%
  \bibfield{journal}{%
  \bibinfo {journal} {Chemical Reviews}\ }%
  \textbf{\bibinfo {volume} {94}},\ \bibinfo {pages} {157} (\bibinfo {year}
  {1994})%
  \bibAnnoteFile{NoStop}{Dha94}%
\bibitem{Voe95}%
  \BibitemOpen
  \bibfield{author}{%
  \bibinfo {author} {\bibfnamefont{P.}~\bibnamefont{Voehringer}}\ and\ \bibinfo
  {author} {\bibfnamefont{N.~F.}\ \bibnamefont{Scherer}},\ }%
  \bibfield{journal}{%
  \bibinfo {journal} {The Journal of Physical Chemistry}\ }%
  \textbf{\bibinfo {volume} {99}},\ \bibinfo {pages} {2684} (\bibinfo {year}
  {1995})%
  \bibAnnoteFile{NoStop}{Voe95}%
\bibitem{Kub62}%
  \BibitemOpen
  \bibfield{author}{%
  \bibinfo {author} {\bibfnamefont{R.}~\bibnamefont{Kubo}},\ }%
  \enquote{\bibinfo {title} {{Fluctuations, Relaxation and Resonance in
  Magnetic Systems}},}\ \ (\bibinfo {publisher} {Oliver and Boyd},\ \bibinfo
  {address} {Edingburgh},\ \bibinfo {year} {1962})\ p.~\bibinfo {pages} {23}%
  \bibAnnoteFile{NoStop}{Kub62}%
\bibitem{Kub63}%
  \BibitemOpen
  \bibfield{author}{%
  \bibinfo {author} {\bibfnamefont{R.}~\bibnamefont{Kubo}},\ }%
  \bibfield{journal}{%
  \bibinfo {journal} {Journal of Mathematical Physics}\ }%
  \textbf{\bibinfo {volume} {4}},\ \bibinfo {pages} {174} (\bibinfo {year}
  {1963})%
  \bibAnnoteFile{NoStop}{Kub63}%
\bibitem{Gam95}%
  \BibitemOpen
  \bibfield{author}{%
  \bibinfo {author} {\bibfnamefont{D.}~\bibnamefont{Gamliel}}\ and\ \bibinfo
  {author} {\bibfnamefont{H.}~\bibnamefont{Levanon}},\ }%
  \emph{\bibinfo {title} {Stochastic Processes in Magnetic Resonance}}\
  (\bibinfo {publisher} {World Scientific Publishing Company Incorporated},\
  \bibinfo {year} {1995})%
  \bibAnnoteFile{NoStop}{Gam95}%
\bibitem{Tan06}%
  \BibitemOpen
  \bibfield{author}{%
  \bibinfo {author} {\bibfnamefont{Y.}~\bibnamefont{Tanimura}},\ }%
  \bibfield{journal}{%
  \bibinfo {journal} {Journal of the Physical Society of Japan}\ }%
  \textbf{\bibinfo {volume} {75}},\ \bibinfo {pages} {082001} (\bibinfo {year}
  {2006})%
  \bibAnnoteFile{NoStop}{Tan06}%
\bibitem{San06}%
  \BibitemOpen
  \bibfield{author}{%
  \bibinfo {author} {\bibfnamefont{F.}~\bibnamefont{Sanda}}\ and\ \bibinfo
  {author} {\bibfnamefont{S.}~\bibnamefont{Mukamel}},\ }%
  \bibfield{journal}{%
  \bibinfo {journal} {The Journal of Chemical Physics}\ }%
  \textbf{\bibinfo {volume} {125}},\ \bibinfo {eid} {014507} (\bibinfo {year}
  {2006})%
  \bibAnnoteFile{NoStop}{San06}%
\bibitem{Har08}%
  \BibitemOpen
  \bibfield{author}{%
  \bibinfo {author} {\bibfnamefont{U.}~\bibnamefont{Harbola}}\ and\ \bibinfo
  {author} {\bibfnamefont{S.}~\bibnamefont{Mukamel}},\ }%
  \bibfield{journal}{%
  \bibinfo {journal} {Physics Reports}\ }%
  \textbf{\bibinfo {volume} {465}},\ \bibinfo {pages} {191 } (\bibinfo {year}
  {2008})%
  \bibAnnoteFile{NoStop}{Har08}%
\bibitem{Dor12}%
  \BibitemOpen
  \bibfield{author}{%
  \bibinfo {author} {\bibfnamefont{K.~E.}\ \bibnamefont{Dorfman}}\ and\
  \bibinfo {author} {\bibfnamefont{S.}~\bibnamefont{Mukamel}},\ }%
  \bibfield{journal}{%
  \bibinfo {journal} {Phys. Rev. A}\ }%
  \textbf{\bibinfo {volume} {86}},\ \bibinfo {pages} {013810} (\bibinfo {year}
  {2012})%
  \bibAnnoteFile{NoStop}{Dor12}%
\bibitem{Sto94}%
  \BibitemOpen
  \bibfield{author}{%
  \bibinfo {author} {\bibfnamefont{H.}~\bibnamefont{Stolz}},\ }%
  \emph{\bibinfo {title} {Time-Resolved Light Scattering from Excitons}},\
  \bibinfo {series} {Springer tracts in modern physics}\ No.\ \bibinfo {number}
  {130}\ (\bibinfo {publisher} {Springer-Verlag},\ \bibinfo {year} {1994})%
  \bibAnnoteFile{NoStop}{Sto94}%
\bibitem{Dor121}%
  \BibitemOpen
  \bibfield{author}{%
  \bibinfo {author} {\bibfnamefont{K.~E.}\ \bibnamefont{Dorfman}}\ and\
  \bibinfo {author} {\bibfnamefont{S.}~\bibnamefont{Mukamel}},\ }%
  \bibfield{journal}{%
  \bibinfo {journal} {Phys. Rev. A}\ }%
  \textbf{\bibinfo {volume} {86}},\ \bibinfo {pages} {023805} (\bibinfo {year}
  {2012})%
  \bibAnnoteFile{NoStop}{Dor121}%
\bibitem{Pol10}%
  \BibitemOpen
  \bibfield{author}{%
  \bibinfo {author} {\bibfnamefont{D.}~\bibnamefont{Polli}}, \bibinfo {author}
  {\bibfnamefont{D.}~\bibnamefont{Brida}}, \bibinfo {author}
  {\bibfnamefont{S.}~\bibnamefont{Mukamel}}, \bibinfo {author}
  {\bibfnamefont{G.}~\bibnamefont{Lanzani}},\ and\ \bibinfo {author}
  {\bibfnamefont{G.}~\bibnamefont{Cerullo}},\ }%
  \bibfield{journal}{%
  \bibinfo {journal} {Phys. Rev. A}\ }%
  \textbf{\bibinfo {volume} {82}},\ \bibinfo {pages} {053809} (\bibinfo {year}
  {2010})%
  \bibAnnoteFile{NoStop}{Pol10}%
\end{thebibliography}

\end{document}